\documentclass[twocolumn]{aastex63}

\received{2023 July 31}
\revised{2023 August 29}
\accepted{2023 September 1}
\published{2023 October 24}
\submitjournal{AJ}

\shorttitle{Simultaneous dual-site technosignature search at 110-190 MHz}
\shortauthors{Johnson et al.}

\usepackage{verbatim}
\usepackage{url}
\usepackage{graphicx}
\usepackage{amsmath}
\usepackage{cleveref}
\usepackage{tikz}
\usepackage{lineno}

\usetikzlibrary{shapes.geometric, arrows}

\tikzstyle{startstop} = [rectangle, rounded corners, minimum width=3cm, minimum height=1cm,text centered, draw=black, fill=red!30]
\tikzstyle{io} = [trapezium, trapezium left angle=70, trapezium right angle=110, minimum width=3cm, minimum height=1cm, text centered, draw=black, fill=blue!30]
\tikzstyle{process} = [rectangle, minimum width=3cm, minimum height=1cm, text centered, draw=black, fill=orange!30]
\tikzstyle{decision} = [diamond, aspect=2, minimum width=3cm, minimum height=1cm, text centered, draw=black, fill=green!30]
\tikzstyle{arrow} = [thick,->,>=stealth]

\newcommand{\surveytargets}{1,631,198 }
\newcommand{\gaiatargets}{1,631,154 }

\newcommand{\distance}{1270 }

\Crefname{figure}{{Figure}}{{Figures}}
\Crefname{equation}{{Equation}}{{Equation}}

\begin{document}

\title{A Simultaneous dual-site technosignature search using international LOFAR stations}
\author[0000-0002-5927-0481]{Owen A. Johnson}
\affiliation{School of Physics, Trinity College Dublin, College Green, Dublin 2, Ireland}
\affiliation{Breakthrough Listen, University of California, Berkeley, 501 Campbell Hall 3411, Berkeley, CA, 94720, USA}
\affiliation{School of Physics, O’Brien Centre for Science North, University College Dublin, Belfield, Dublin 4, Ireland}
\author[0000-0002-8604-106X]{Vishal Gajjar}
\affiliation{SETI Institute, 339 Bernardo Ave Suite 200 Mountain View, CA 94043, USA}
\affiliation{Breakthrough Listen, University of California, Berkeley, 501 Campbell Hall 3411, Berkeley, CA, 94720, USA}
\author[0000-0002-4553-655X]{Evan F. Keane}
\affiliation{School of Physics, Trinity College Dublin, College Green, Dublin 2, Ireland}
\affiliation{School of Natural Sciences, Ollscoil na Gaillimhe --- University of Galway, University Road, H91 TK33, Galway, Ireland}
\author[0000-0001-7185-1310]{David J. McKenna}
\affiliation{School of Physics, Trinity College Dublin, College Green, Dublin 2, Ireland}
\affiliation{School of Cosmic Physics, Dublin Institute for Advanced Studies, 31 Fitzwilliam Place, Dublin 2, Ireland}
\author[0000-0003-1517-8891]{Charles Giese}
\affiliation{Max-Planck-Institut für Radioastronomie, Bonn, Germany}
\affiliation{School of Physics, Trinity College Dublin, College Green, Dublin 2, Ireland}
\author[0000-0001-7060-4852]{Ben McKeon}
\affiliation{School of Natural Sciences, Ollscoil na Gaillimhe --- University of Galway, University Road, H91 TK33, Galway, Ireland}
\affiliation{Department of Physics, University of Limerick, V94 T9PX, Limerick, Ireland}
\author[0000-0002-4963-179X]{Tobia D.\ Carozzi}
\affiliation{Department of Space, Earth and Environment, Chalmers University of Technology, Onsala Space Observatory, 439 92 Onsala, Sweden}
\author[0000-0001-6299-5726]{Cloe Alcaria}
\affiliation{School of Physics, Trinity College Dublin, College Green, Dublin 2, Ireland}
\affiliation{Département de Physique, Université Paul Sabatier, Toulouse, France}
\author[0000-0002-7050-0161]{Aoife Brennan}
\affiliation{School of Physics, Trinity College Dublin, College Green, Dublin 2, Ireland}
\affiliation{School of Natural Sciences, Ollscoil na Gaillimhe --- University of Galway, University Road, H91 TK33, Galway, Ireland}
\affiliation{European Southern Observatory, Alonso de Córdova 3107, Vitacura, Región Metropolitana, Chile}
\author[0000-0002-7461-107X]{Bryan Brzycki}
\affiliation{Breakthrough Listen, University of California, Berkeley, 501 Campbell Hall 3411, Berkeley, CA, 94720, USA}
\author[0000-0003-4823-129X]{Steve Croft}
\affiliation{Breakthrough Listen, University of California, Berkeley, 501 Campbell Hall 3411, Berkeley, CA, 94720, USA}
\affiliation{SETI Institute, 339 Bernardo Ave Suite 200 Mountain View, CA 94043, USA}

\author{Jamie Drew}
\affiliation{The Breakthrough Initiatives, NASA Research Park, Bld. 18, Moffett Field, CA, 94035, USA}

\author{Richard Elkins}
\affiliation{Breakthrough Listen, University of California, Berkeley, 501 Campbell Hall 3411, Berkeley, CA, 94720, USA}

\author[0000-0001-9745-0400]{Peter T. Gallagher}
\affiliation{School of Cosmic Physics, Dublin Institute for Advanced Studies, 31 Fitzwilliam Place, Dublin 2, Ireland}
\author[0000-0002-5004-3573]{Ruth Kelly}
\affiliation{Mullard Space Science Laboratory, Department of Space and Climate Physics, UCL,  Surrey RH5 6NT, United Kingdom}
\author[0000-0002-7042-7566]{Matt Lebofsky}
\affiliation{Breakthrough Listen, University of California, Berkeley, 501 Campbell Hall 3411, Berkeley, CA, 94720, USA}
\author[0000-0001-6950-5072]{Dave H.E. MacMahon}
\affiliation{Breakthrough Listen, University of California, Berkeley, 501 Campbell Hall 3411, Berkeley, CA, 94720, USA}
\author[0000-0003-4399-2233]{Joseph McCauley}
\affiliation{School of Physics, Trinity College Dublin, College Green, Dublin 2, Ireland}
\author[0000-0002-4278-3168]{Imke de Pater}
\affiliation{Breakthrough Listen, University of California, Berkeley, 501 Campbell Hall 3411, Berkeley, CA, 94720, USA}
\author[0000-0002-0598-2061]{Shauna Rose Raeside}
\affiliation{School of Physics, Trinity College Dublin, College Green, Dublin 2, Ireland}
\affiliation{School of Cosmic Physics, Dublin Institute for Advanced Studies, 31 Fitzwilliam Place, Dublin 2, Ireland}
\affiliation{School of Physical Sciences and Centre for Astrophysics \& Relativity, Dublin City University, Glasnevin, D09 W6Y4, Ireland}
\author[0000-0003-2828-7720]{Andrew P.V. Siemion}
\affiliation{Breakthrough Listen, University of California, Berkeley, 501 Campbell Hall 3411, Berkeley, CA, 94720, USA}
\affiliation{SETI Institute, 339 Bernardo Ave Suite 200 Mountain View, CA 94043, USA}
\author{S. Pete Worden}
\affiliation{The Breakthrough Initiatives, NASA Research Park, Bld. 18, Moffett Field, CA, 94035, USA}

\begin{abstract}
    The Search for Extraterrestrial Intelligence (SETI) aims to find evidence of technosignatures, which can point towards the possible existence of technologically advanced extraterrestrial life. Radio signals similar to those engineered on Earth may be transmitted by other civilizations, motivating technosignature searches across the entire radio spectrum. In this endeavor, the low-frequency radio band has remained largely unexplored; with prior radio searches primarily above $1$~GHz. In this survey at $110-190$~MHz, observations of \surveytargets{} targets from \textit{TESS} and \textit{Gaia} are reported. Observations took place simultaneously with two international stations (non-interferometric) of the Low Frequency Array in Ireland and Sweden. We can reject the presence of any Doppler drifting narrow-band transmissions in the barycentric frame of reference, with equivalent isotropic radiated power of $10^{17}$~W, for 0.4 million (or 1.3 million) stellar systems at $110$ (or $190$)~MHz. This work demonstrates the effectiveness of using multi-site simultaneous observations for rejecting anthropogenic signals in the search for technosignatures. 
\end{abstract}


\section{Introduction}
\label{sect:intro}
In the last 50 years, evidence has steadily mounted, that the constituents and conditions necessary for life are common in the Universe \citep{Wordsworth2014}. Predicting specific properties of electromagnetic emissions from extraterrestrial technologies is one of the most challenging aspects of searching for life in the universe. However, it also represents a high-risk, high-reward endeavor. If an extraterrestrial civilization were intentionally attempting to indicate its presence through such emissions, it would be advantageous to make the signals easily distinguishable from natural phenomena. The evidence of such emissions is referred to as `technosignatures', and the field dedicated to their detection is known as the Search for Extra-terrestrial Intelligence (SETI).

It is commonly assumed that civilizations elsewhere in the universe may employ similar technologies to those developed on Earth. Consequently, radio frequencies are considered a logical domain for conducting SETI surveys due to the widespread use of telecommunications and radar. Therefore, radio astronomy has played a significant role in the field of SETI since the 1960s \citep{Drake:1961bv, Tarter:1980p1516}. Numerous previous SETI surveys have utilized large single dish telescopes operating at frequencies $\gtrsim$ 1\,GHz \citep{Tarter:1996jf, Siemion_KEPLER_ApJ, Enriquez:2017} \footnote{See \citealt{wright18} for a review}. However, exploration of the radio window below 1\,GHz has been relatively limited. Technosignature searches commonly seek narrowband (approximately Hz-scale) radio emissions, either transmitted directly or leaking from other civilizations. Nonetheless, there is no inherent preference for any specific segment of the radio spectrum, which necessitates surveys spanning from low frequencies (30\,MHz) to high frequencies (100\,GHz; \citealt{Cherry_2022}). At 30\,MHz it becomes very difficult to observe from the ground due to ionospheric conditions \citep[see][chap.~7.8]{burke_graham-smith_wilkinson_2019}. This study primarily focuses on low-frequency SETI in the 110 - 190\,MHz range.

\subsection{Scientific Motivation}
Low-frequency radio SETI presents significant challenges due to higher sky temperatures, which limit the sensitivity of the underlying observations. The Murchison Widefield Array (MWA, \citealt{2013PASA...30....7T}) in Western Australia has been at the forefront of low-frequency SETI research thus far \citep{tingay:center,tingay:anticenter,tingay:omm}. However, the LOw Frequency ARray (LOFAR) presents a compelling scientific case for conducting a SETI survey \citep{Eavesdropping}. Aside from operating at low frequencies in the northern sky, LOFAR offers a large field of view, enabling the search for technosignatures across thousands of stars in each observation. 

Radio SETI also grapples with the challenge of handling a significant amount of radio frequency interference (RFI). Traditionally, SETI surveys have been conducted using single dish radio telescopes. While these telescopes offer operational convenience and room for upgrades, they possess limitations in effectively distinguishing between sources of interference and authentic sky-bound signals unless equipped with multibeam receivers. In contrast, the utilization of two local LOFAR stations presents two notable advantages over conventional single dish surveys. As demonstrated in studies conducted by \cite{Enriquez:2017} and \cite{danny-parkes}, single dish surveys typically employ an 'ON' and 'OFF' observing technique, where the target is observed for five minutes ('ON') followed by five minutes of observing a different location ('OFF'). This cycle is repeated three times, resulting in three 'ON' pointings and three 'OFF' pointings. This approach facilitates the identification and elimination of narrowband signals detected in the local environment that could potentially interfere with the search for technosignature candidates. By employing multiple stations, the search benefits from the unique local RFI environments at each station. This leads to a higher rate of rejecting false positive signals compared to the aforementioned surveys, which have signals of interest on the order of thousands. Additionally, since there are two stations involved, there is no requirement to alternate between an 'ON' and 'OFF' observation regime. As a result, the entire observation duration can be dedicated to directly observing the target, as the comparison of RFI environments would yield the same effect. This characteristic renders SETI surveys with two or more telescopes a highly valuable resource, particularly in today's RFI environments.

\subsection{Breakthrough Listen}
The Breakthrough Listen (BL) program is conducting one of the most comprehensive searches for evidence of intelligent life by extending the search to a wide variety of targets from existing ground-based observing facilities (see \citealt{gsc+19} for a review). All of the existing observations within the BL program have so far been conducted in the $1-27.45$~GHz range \citep{Cherry_2022}. \\ 
In this paper we report on low frequency extension of the BL initiative using two international LOFAR stations to perform simultaneous dual-site observations of nearby exoplanet candidates of interest from both the Transiting Exoplanet Survey Satellite \citep[TESS;][]{TESS_2015} and from \textit{Gaia} \citep{Gaia_DR2_HR_2018} in collaboration with the BL program. This survey also demonstrates the proof of concept of using dual-site observations for the rejection of spurious local sources of terrestrial origin. This method thus removes the need for separate `ON' and `OFF' pointings \citep{Gajjar_2021_BLGC1}. In \S~\ref{sec:observations} we describe the observational set up and the data acquired. \S~\ref{sect:signal_types} explains the data analysis steps taken. We discuss the implications of this work in \S~\ref{sec:discussion} before concluding in \S~\ref{sec:conclusion}.
\section{Observations}\label{sec:observations}
\begin{figure}[h!]
    \centering
    \includegraphics[width = 0.5\textwidth]{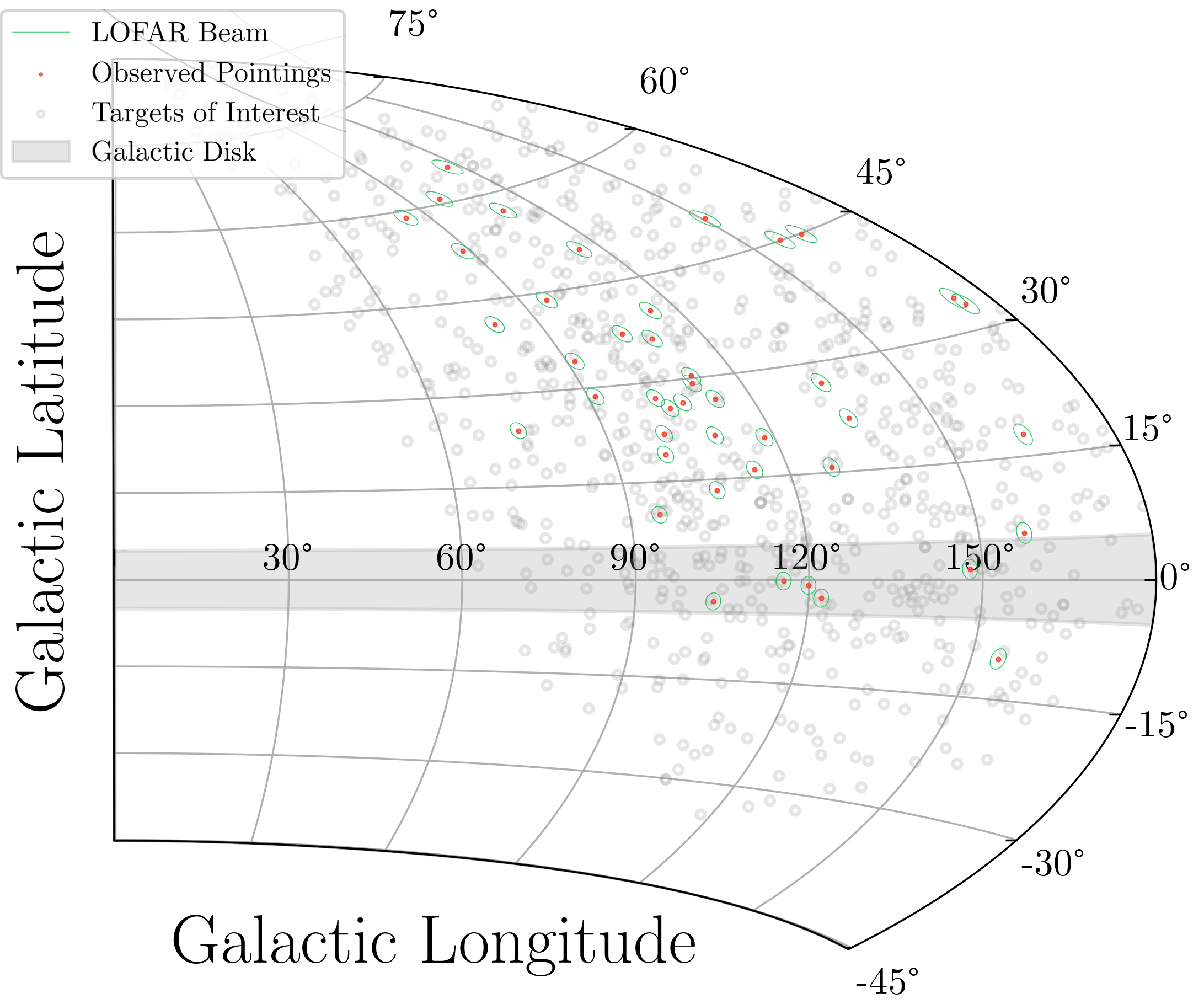}
    \caption{An Aitoff projection of the sky in Galactic coordinates depicts the distribution of survey pointings, with the Galactic disk shaded in grey. \textit{Gaia} sources are omitted from the plot due to their extremely high source density. Grey dots represent \textit{TESS} Targets of Interest (ToIs). Those targets observed during our survey are marked with red dots at boresight with green showing the half-power beam width ($2.59^\circ$).
  }
  \label{fig:TESS_TOIs}
\end{figure}

This study encompasses a total of $44$ targeted pointings, where each pointing consists of a 15-minute scan centered on specific targets selected from the \textit{TESS} catalog, focusing on confirmed or candidate exoplanets (refer to Figure \ref{fig:TESS_TOIs}). The entire observation campaign spanned a duration of 11 hours, covering an area of 232 deg$^2$ in the northern sky. 

\subsection{LOFAR}
LOFAR, a pioneering low-frequency aperture array telescope, spans hundreds of kilometers across Europe and serves as a pathfinder to the Square Kilometer Array (SKA). The array consists of a core station with outrigger stations situated in the Netherlands and additional international stations spanning multiple countries, such as Germany, France, Sweden, Ireland, Latvia, Poland, and the United Kingdom. Additionally, stations are currently in the process of being constructed in Italy and Bulgaria. The LOFAR array operates using two types of antenna, the Low Band Antenna (LBA) and the High Band Antenna (HBA), operating at 10-90 MHz and 100-250 MHz respectively. In this study, the HBAs at the Irish and Swedish LOFAR station are used to carry out observations non-interferometrically.  The field-of-view (FoV) of an international LOFAR station is rather large; at full-width-half maximum it is $5.3$, $3.4$ and $2.3$ deg$^2$ at frequencies of $120$, $150$ and $180$~MHz, respectively~\citep{2013arXiv1305.3550V}. With such a large region where our observations are sensitive, each pointing contains millions of stars that can be searched for technosignatures \citep{Bart-Wlodarczyk-Sroka}. In this survey, in addition to the $44$ \textit{TESS} targets at boresight, \gaiatargets \textit{Gaia} targets are covered by our observations and so are searched for technosignatures. 

 \subsection{Targets} \label{sc:targets}

 A significant fraction of radio emission from Earth is emitted in the direction of the ecliptic plane. For example, powerful planetary radars are used to explore solar system objects \citep{Siemion_KEPLER_ApJ} and high-powered transmitters are used to communicate with solar system probes \citep{Enriquez:2017}. It is conceivable then that such leakage radiation may also be emanating from other worlds, preferentially in \textit{their} planetary orbital planes. This is why we chose \textit{TESS} targets, as these are the closest transiting exoplanet systems known \citep{kepler_2008,TESS_2015}. Observing these sources with the LOFAR HBAs enables robust constraints on any associated artificial low-frequency radio emission.

\subsubsection{\textit{TESS} Targets} In order to determine a target list for this work, the latest list of \textit{TESS} object of interests (TOIs) was retrieved \citep{NEA12} and a shortlist of targets were obtained rejecting possible false positives. Since the sensitivity of the HBA array is best $\pm$30$^\circ$ of zenith, the required overlap for both LOFAR international stations spans the declination range +27$^\circ$ to +83$^\circ$. Further practical considerations were also accounted for, i.e. to stay as far from bright sources like the Sun as possible at both sites, and to balance sensitivity at both sites \citep{D-SOP}\footnote{These selection criteria were applied using a custom developed dual-site observation planning software.}. We report observations towards 44 unique targets from the \textit{TESS} catalogue in this study, where each target was observed for 15 minutes. Figure \ref{fig:TESS_TOIs} shows the distribution of these targets observed in comparison to the pool of all \textit{TESS} TOIs. 

\subsubsection{Gaia Sources}
\begin{figure}
    \centering
    \includegraphics[width = 0.45\textwidth]{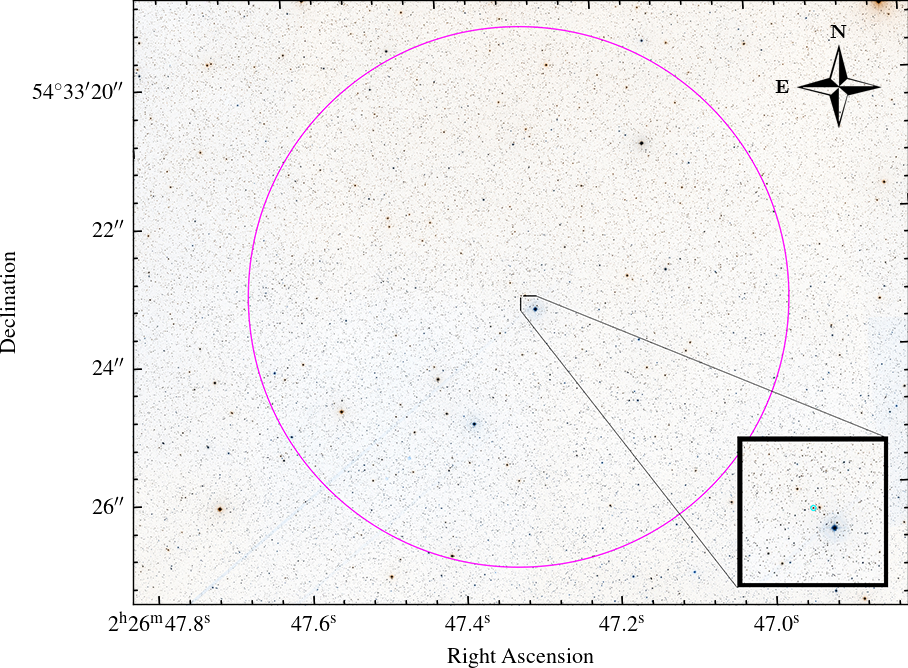}
    \caption{This figure illustrates one of the boresight pointings directed at a \textit{TESS} object (highlighted in green), with the LOFAR beam's Full-Width Half-Maximum (FWHM) shown in pink. The background image is from the Sloan Digital Sky Survey \citep[SDSS;]{SDSSDR13}.}
    \label{fig:inbeam_trgts_SDSS}
\end{figure}

The beam of a LOFAR station has an expansive coverage enabling observation of a substantial number of stars in the field-of-view. The significance of these in-field stars has been highlighted by \cite{Bart-Wlodarczyk-Sroka}. Consequently, during our observations targeting 44 sources from the TESS catalogue, we encountered a significant number of in-field stars within our field-of-view as shown in Figure \ref{fig:inbeam_trgts_SDSS}. 

To determine the list of targets within this field-of-view, we utilized the \textit{Gaia} catalogue. Some previous major SETI surveys focused their searches towards Sun-like stars \citep{Tarter:1996jf}. However, since our understanding of the origin of life is limited, it makes sense to allow for the possibility of life arising on a planet that is neither Earth-like, nor around stars that are Sun-like. Similarly, we should consider planets not necessarily located in the habitable zone. This is typically characterized as the orbital range wherein liquid water could exist~\citep{KASTING1993108}, as inferred from planetary equilibrium temperatures often ignoring the unknown albedo of the exoplanets. Any sensitive radio SETI survey seeking to maximize the chance of detecting weak radio signals should, insofar as possible, expand its search to encompass nearby stars of a broad range of spectral types and with exoplanets of all sizes and distances from their parent star. Thus, we conducted calculations to determine the number of \textit{Gaia} stars with a mean distance of 1215 pc, with an accuracy in their distances of at least 20\%. \\

\begin{figure}
    \centering
    \includegraphics[width = 7cm]{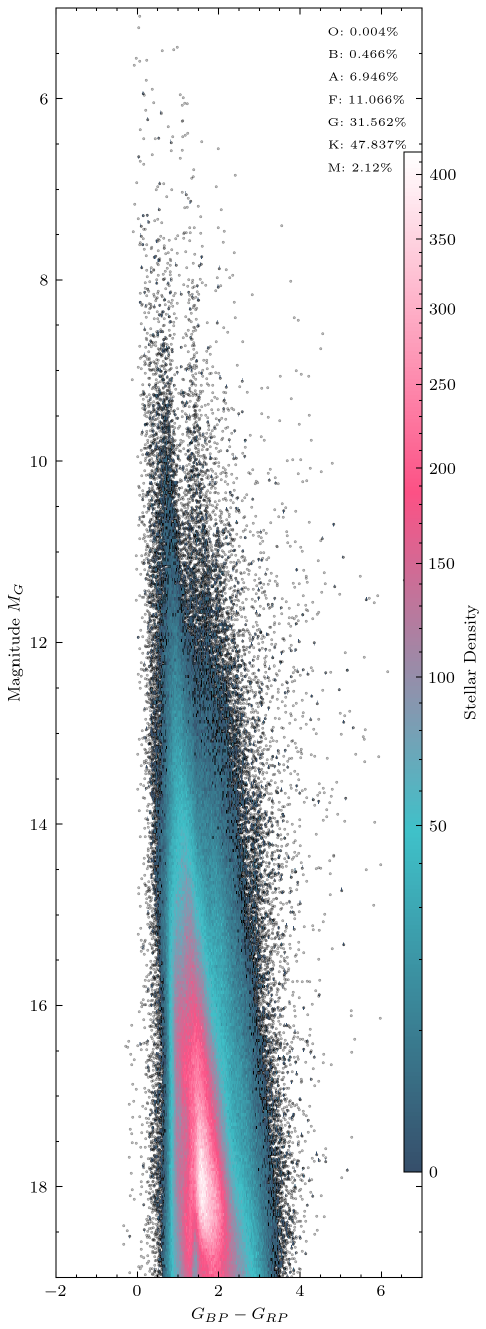}
    \caption{A Hertzsprung-Russell diagram of the \surveytargets \textit{Gaia} targets searched for technosignatures in this survey with a mean distance of \distance pc. The relative distribution with respect to spectral type is: O $<0.01\%$, B---$0.4\%$, A---6.94\%, F---11.01\%, G---31.56\%, K---47.84\%, M---2.12\%. This is as opposed to the general \textit{Gaia} catalogue which is: O, B $<0.0001\%$, A---1.5\%, F---18.9\%, G---44.4\%, K---34.5\%, M---0.6\%.}
    \label{fig:HR-Diagram}
\end{figure}

This study used \textit{Gaia}'s third data release  \citep[GDR3;][]{2023GDR3, astroquery}. When analyzing GDR3 two filters were applied to the in beam target values survey volume and sensitivity accuracy. Firstly a constraint on the RA and DEC errors were implemented. If a \textit{Gaia} source was found to be in the beam but had a error magnitude greater than the FWHM it was removed from the source pool. \Cref{eq:condition1} states the first condition of filtering. 
\begin{equation}
    \theta_\text{sep} + \sqrt{\Delta \text{RA}^2 + \Delta \text{DEC}^2} < 1.295^\circ 
    \label{eq:condition1}
\end{equation}
As the sensitivity of the survey is calculated based on a source's distance a second filter is implemented to remove sources that have large errors. By taking the difference $(\Delta \sigma_G)$ in the upper and lower confidence levels of GSP-Photometry\footnote{SQL Keys: \texttt{distance\_gspphot}, \texttt{distance\_gspphot\_upper} $(d_{M_G})$ and \texttt{distance\_gspphot\_lower}} to obtain a percentage error on distance. All sources with a $d_{M_G}$ error of 20\% or greater are filtered out of the source list. Equation (\ref{eq:condition2}) states the second condition of filtering:
\begin{equation}
    \dfrac{\Delta \sigma_G}{d_{M_G}} < 20\% 
    \label{eq:condition2}
\end{equation}
A total of $\gaiatargets$ stars from this list, making it one of the largest samples of stars ever surveyed for SETI purposes. Figure~\ref{fig:HR-Diagram} shows a Hertzsprung-Russell diagram for the targets. 

\begin{figure}
    \centering
    \includegraphics[width = 0.45\textwidth]{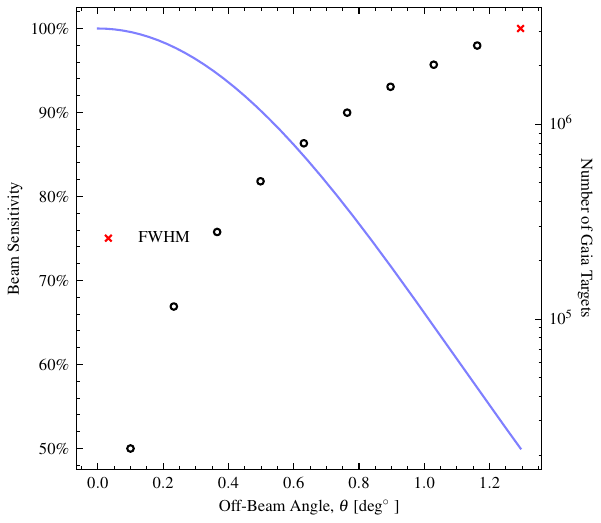}
    \caption{The LOFAR HBA beam's sensitivity at 150~MHz changes in relation to the off-axis bore-sight angle ($\theta$). Furthermore, a plot of the number of filtered \textit{Gaia} targets within the beam pointing as a function of $\theta$ is presented. The \textcolor{red}{$\times$} represents the value for the Full Width Half Max of the beam.}
    \label{fig:off_beam_sens}
\end{figure}

\subsection{Simultaneous observations}

\begin{figure*}
    \centering
    \includegraphics[width = 0.78\textwidth]{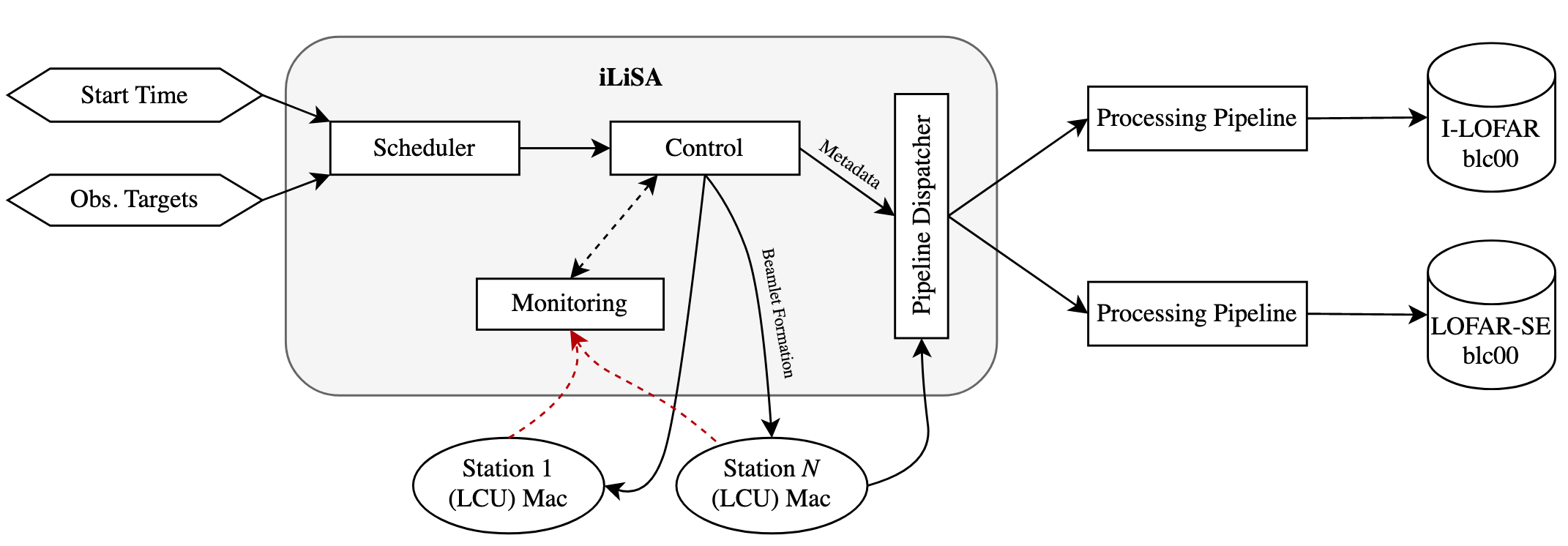}
    \caption{\texttt{iLiSA} block diagram, showing how an operator-defined schedule is ingested and is executed in a timely fashion. Data then begins to flow and ends up being ingested by processing pipelines together with associated metadata.} 
    \label{fig:iLiSA}
\end{figure*}

Typically, international LOFAR stations operate as standalone telescopes $2-3$ days per week, i.e. they do not operate as part of the International LOFAR Telescope's Europe-wide array. This project was undertaken during this standalone time. For this purpose the \emph{international LOFAR in Stand-Alone mode} (\texttt{iLiSA}) package\footnote{\url{https://github.com/2baOrNot2ba/iLiSA/releases/tag/v6.1}} is used to control both telescopes simultaneously. \texttt{iLiSA} provides a high-level operational control of multiple LOFAR stations, including scheduling, processing pipeline dispatching and metadata aggregation (see Figure~\ref{fig:iLiSA}). For the observations in this study, an operator-produced list of targets that were close\footnote{In practice as Birr and Onsala are separated by $\sim20\deg$ of longitude, the optimum scheduling is to observe $\sim 40$~min$-T_{\rm obs}/2$ `late' at Onsala and $40$~min$+T_{\rm obs}/2$ `early' at Birr.} to the local meridian was fed into \texttt{iLiSA} at each epoch. Towards each target, one beam per station was formed using \texttt{iLiSA}, and each beam was formed with $412$ HBA sub-bands (corresponding to a bandwidth of $80.46875$~MHz). The scan time on each target was $15$~min, and the whole scheduling block was a few hours per epoch.

In this study, the only \texttt{iLiSA} pipeline used was the raw data recorder, which simply receives UDP packets of beam-formed data and writes them to disk, separately at each site. The data consist of coarse-channelised complex voltages. The initial data stream is two polarization's from each antenna. With real Nyquist-sampling with a $200$-MHz clock, with a coarse channelisation factor of $512$ taking place, only $488$ ($244$) of these being recordable when the data are written as 8-bit (16-bit) complex numbers. Further processing of these data is necessary and we use \texttt{udpPacketManager}~\citep{David_JOSS} to ultimately create total intensity  \texttt{sigproc} \citep{lorimerSIGPROCPulsarSignal2011} formatted filterbank files at the time and frequency resolution appropriate for our Doppler-drift search, shown in Table~\ref{tab:data_details}.

\begin{deluxetable}{lc}
  \tablecaption{\label{tab:data_details}Specifications for the data input to our processing pipeline after pre-processing and preparing the raw data with \texttt{udpPacketManager}~\citep{David_JOSS}.}
  \tablewidth{0pt}
  \tablehead{
  \colhead{Data Attribute} & \colhead{Value}
  }  
  \startdata
    Frequency Start $(f_\text{start})$ & 109.9609375 MHz\\ 
    Frequency End $(f_\text{end})$ & 190.0390625 MHz \\
    Frequency Resolution & $2.980232239$~Hz \\
    Channel Number & $27000832$ \\
    Obs. Length ($t_{\mathrm{obs}}$) & 15 mins\\
    Temporal Resolution & $0.67108864$~s \\ 
  Maximum Drift Rate Search & $\pm4 \ \text{Hz s}^{-1}$ \\
  $\text{S}/\text{N}_{\text{min}}$ threshold & $10$ \\ 
  \enddata

\end{deluxetable}

\begin{figure}
    \centering
    \includegraphics[width = 0.42\textwidth]{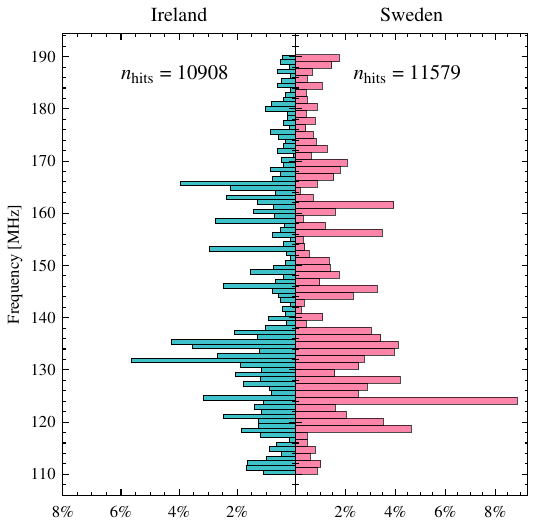}
    \includegraphics[width = 0.42\textwidth]{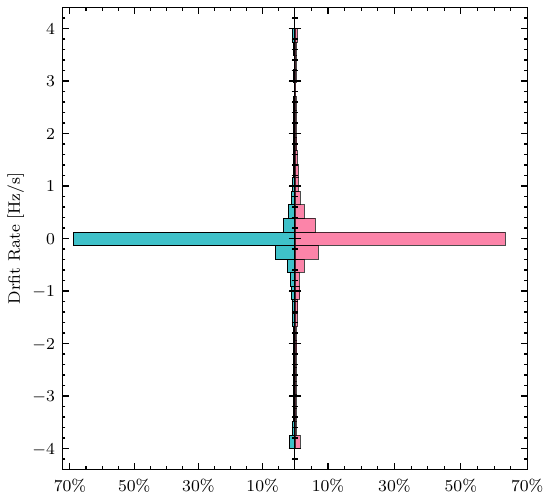}
    \includegraphics[width = 0.42\textwidth]{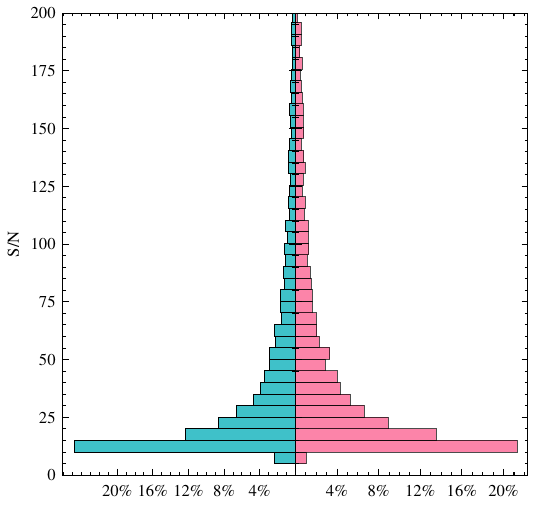}
    \caption{Comparison of drifting signals or `hits' detected at both stations seen across the HBA frequency band. Each bin within the data set represents a 1 MHz frequency range and is accompanied by a corresponding percentage indicating its proportion to the overall data set.}
    \label{fig:histogram_hits_comparison}
\end{figure}

\section{Analysis}
\label{sect:signal_types}
The most common transmissions sought in SETI searches are narrow-band ($\sim$Hz) radio signals. Ubiquitous in early terrestrial communication systems, such signals can be produced with relatively low energy and traverse the interstellar medium easily. They can be readily distinguished from natural astrophysical sources. These signals could either be transmitted intentionally or arise as leakage from extra-solar technologies. The apparent frequency of a distant narrow-band transmitter is expected to exhibit Doppler drift due to the relative motion between the transmitter and receiver. 
The Breakthrough Listen group has developed an efficient narrow-band search software package which includes a search for such drifting signals, named \verb|turboSETI|~\citep{Enriquez:2017} which we use for the analysis of our LOFAR survey data. However, before we can compare narrowband signal hits detected using this tool, it was necessary to compare them in the Barycentric reference frame. 

\begin{figure}
    \centering
    \includegraphics[width = 0.45\textwidth]{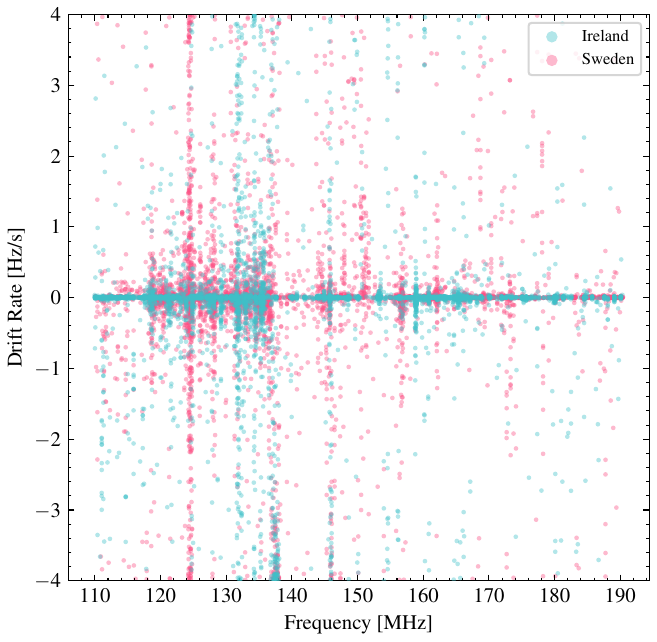}
    \caption{A scatter plot of the drift rate values against detected frequency. The Irish station is shown in pink and the Swedish station is shown in blue.}
    \label{fig:scatter_DR}
\end{figure}

\subsection{Barycentric correction}
The movement of the Earth around its axis and the Sun introduces a Doppler effect that causes radio signals' frequency and arrival time to shift, as 
\begin{equation}
f_{obs} = f_{em} \left( 1 - v_{rel} \right).
\label{eq:doppler}
\end{equation}
Here, $f_{obs}$ is the observed frequency, $f_{em}$ is the emitted frequency (or barycentric frequency), and $v_{rel}$ is the velocity of the source relative to the observer, normalized by the speed of light. The Doppler effect only depends on the velocity of the source relative to the observer, and for a source moving towards the observer, we can consider $-v_{rel}$, while for a source moving away from the observer, we consider $+v_{rel}$. The relative velocity between the observer and the source will differ for different observing epochs, the location of the source in the sky, and the geographical location of the telescope. For instance, the target TIC\,27677846 was observed on UT 2021 July 15, from both the LOFAR stations simultaneously. The expected relative velocity ($v_{rel}$) towards the source was $-7.669\times10^{-5}$ and $-7.496\times10^{-5}$, causing a relative shift ($f_{em}-f_{obs}$) of +11.504\,kHz and +11.246\,kHz for a hypothetical ETI signal transmitted at a constant frequency of 150\,MHz observed at the Sweden and Ireland stations, respectively. These are significant shifts that are distinct at the two stations, and they need to be corrected to compare the same signal observed at the two stations. 

Typically, barycentric corrections are introduced by adjusting the local oscillator during observations. However, in our study, we record beamformed baseband voltages during observations and produce three different data products with varying temporal and spectral resolutions during post-processing \citep{2019_Lebofsky}. We are interested in searching for a wide variety of signals, including narrowband signals, broadband transient signals, and wide-band pulsating signals. Introducing local oscillator shifts during observations can impact our other signal searches. Therefore, we correct for barycentric drift after the channelization and detection of the baseband voltages for narrowband signal searches using software which we have developed\footnote{\url{https://github.com/gajjarv/BaryCentricCorrection}}. Details of this correction and code \citep{Vishal_Bary} are fully discussed in Appendix~\ref{appen:barycenter}. For comparison both topocentric and barycentric data are analysed for a small sub-set of the targets and the results are shown in Table \ref{tab:barycenter_vs_topocenter}.

\begin{deluxetable*}{lCCC|CCC}
  \tablecaption{Comparison of the number narrow-band signals detected for a sub-set of \textit{TESS} objects in both the topocentric and barycentric reference frame. \texttt{turboSETI} searches were completed using identical parameters in each reference frame. We did find a small number of mutual hits while it was found that Barycentrically correcting the data prior to searching filtered out all prior mutual hits of interest. \label{tab:narrowband_hit_bary_non_bary_compare}}
  \label{tab:barycenter_vs_topocenter}
  \tablewidth{0pt}
  \tablehead{
  \colhead{TIC ID} & \colhead{LOFAR-SE Hits} &\colhead{I-LOFAR Hits} & \colhead{Mutual Hits} & \colhead{LOFAR-SE Hits} &\colhead{I-LOFAR Hits} & \colhead{Mutual Hits}\\   
   &  \multicolumn{3}{c}{\emph{Topocentric Results}}  &  \multicolumn{3}{c}{\emph{Barycentrically Corrected Results}} }  
  \startdata
121966220 & 387 & 178 & 18 & 376 & 171 & 0 \\
249862365   & 340 & 193 & 22 & 330 & 196 & 0 \\ 
250724252  & 276 & 190 & 21 & 265 & 262 & 0 \\ 
27677846  & 313 & 159 & 20 & 294 & 151 & 0 \\ 
470315428 & 384  & 175 & 17 & 356 & 224 & 0 \\ 
  \enddata

\end{deluxetable*}

\subsection{Searching for Narrowband Signals}
Using \verb|turboSETI| with the parameters outlined in Table \ref{tab:data_details}, a Doppler drift search was carried out on the observed candidates listed in \Cref{tab:TESS_results} and \ref{tab:Gaia_results} at both stations. This resulted in the list of `hits' collected in Tables \ref{tab:TESS_results} and \ref{tab:Gaia_results}, where hits are defined as a narrow-band signal detected above the given threshold, S/N = 10. The distribution of narrow-band signals detected at both stations is shown in \Cref{fig:histogram_hits_comparison}. A large percentage of hits are seen at both sites in the 120 - 140 MHz range. This falls within the range of expected RFI leakage seen from neighboring airports\footnote{Shannon \& Goteborg Landvetter}. \\

Using a drift-rate search of $\pm 4 \ \text{Hz/s}$ for this study covers a fraction of the possible drift rates of transmitters from exotic objects that can be detected as outlined by \cite{Sheikh_2019}. \cite{LiNarrowBand} shows that 4~Hs/s is comprehensive in relative to the expected distribution of exoplanet drift rates. The omission of a search in this parameter space is due to its computationally intense nature of searching for narrow-band signals across a sizeable drift-rate range. However, in doing this, the parameter space searched for ETI signal has been drastically reduced. Continual development of search algorithms like \verb|turboSETI| is progressing to make larger drift-rates searches a more computationally feasible. \\ 

Upon first inspection of \Cref{fig:histogram_hits_comparison} it appears that the results at both stations are somewhat similar. However, upon performing a Kolmogorov-Smirnov (KS) test for each set of results for drift-rate, SN and frequency of detected hits the highest $p$-value returned was on the order of $10^{-11}$ indicating that the RFI environments at each of the stations are significantly different.

\subsection{Dual-site coincidence rejection}
\begin{figure*}[ht]
    \centering
    \includegraphics[width = \textwidth]{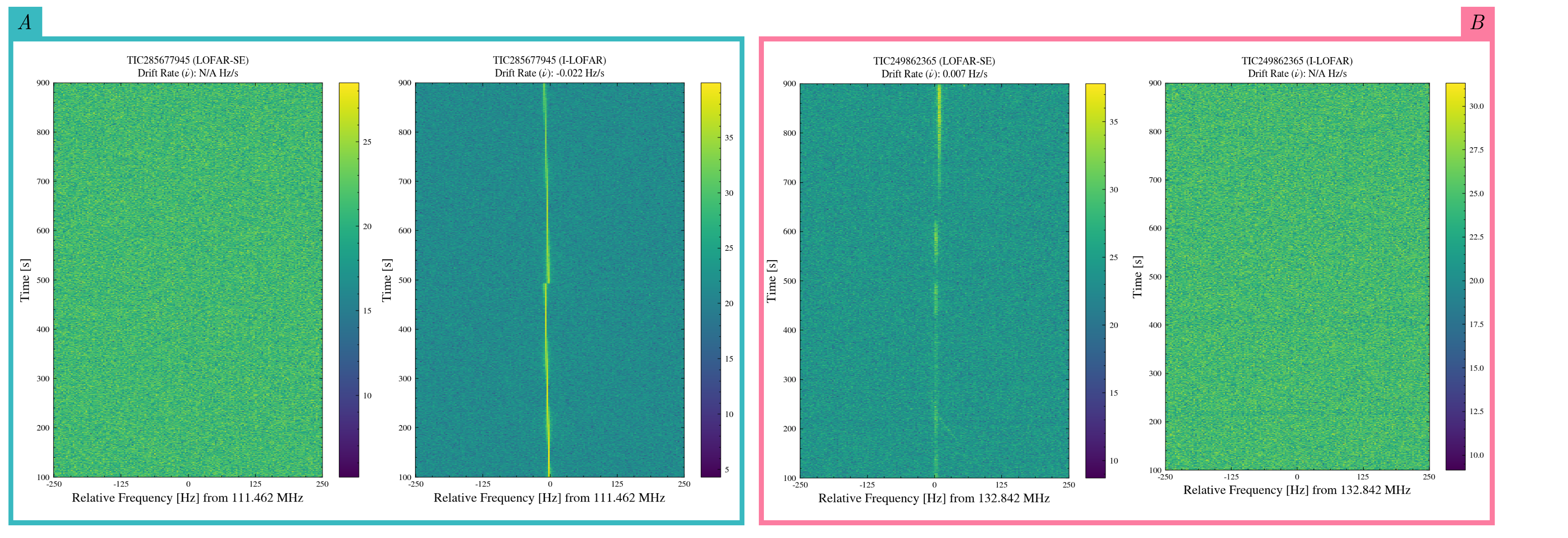}
    \caption{Dynamic spectra (waterfall plots) of detected narrow-band signal centered on the detected frequency of detection, showing the two most common cases of coincidence rejection. Case A (\textit{left}) shows a narrowband signal with a non-zero drift rate detected at I-LOFAR and not LOFAR-SE. Case B \textit{(Right)} shows the opposite where a non-zero drift rate signal is not detected by I-LOFAR but is detected by LOFAR-SE. For a signal to be considered a detection of interest both sites would have to exhibit non-zero drift rate signal at the same frequency simultaneously. }
    \label{fig:coincidence_cases}
\end{figure*}

 In the case of this study, a singular beam observes a single target for 15 minutes at both stations and observations are converted to barycentric reference frame. Narrow-band searches are then performed at both sites, and the results of both searches are compared. In \Cref{fig:coincidence_cases} two common detection cases are shown and how the use of dual site observation aids in the nature of each signals origin. 

\textit{Case A:} In this case a hit has been detected at the Irish station but is absent from the same frequency at the Swedish station. Thus the signal is rejected as a extraterrestrial emitter and deemed as RFI local to the Irish station. \\ 

\textit{Case B:} Similar to case A but this time the converse is seen, a hit has been detected at Swedish station and not at a the frequency at the Irish station. \\ 

\begin{figure*}[]
    \centering
    \includegraphics[width = 0.9\textwidth]{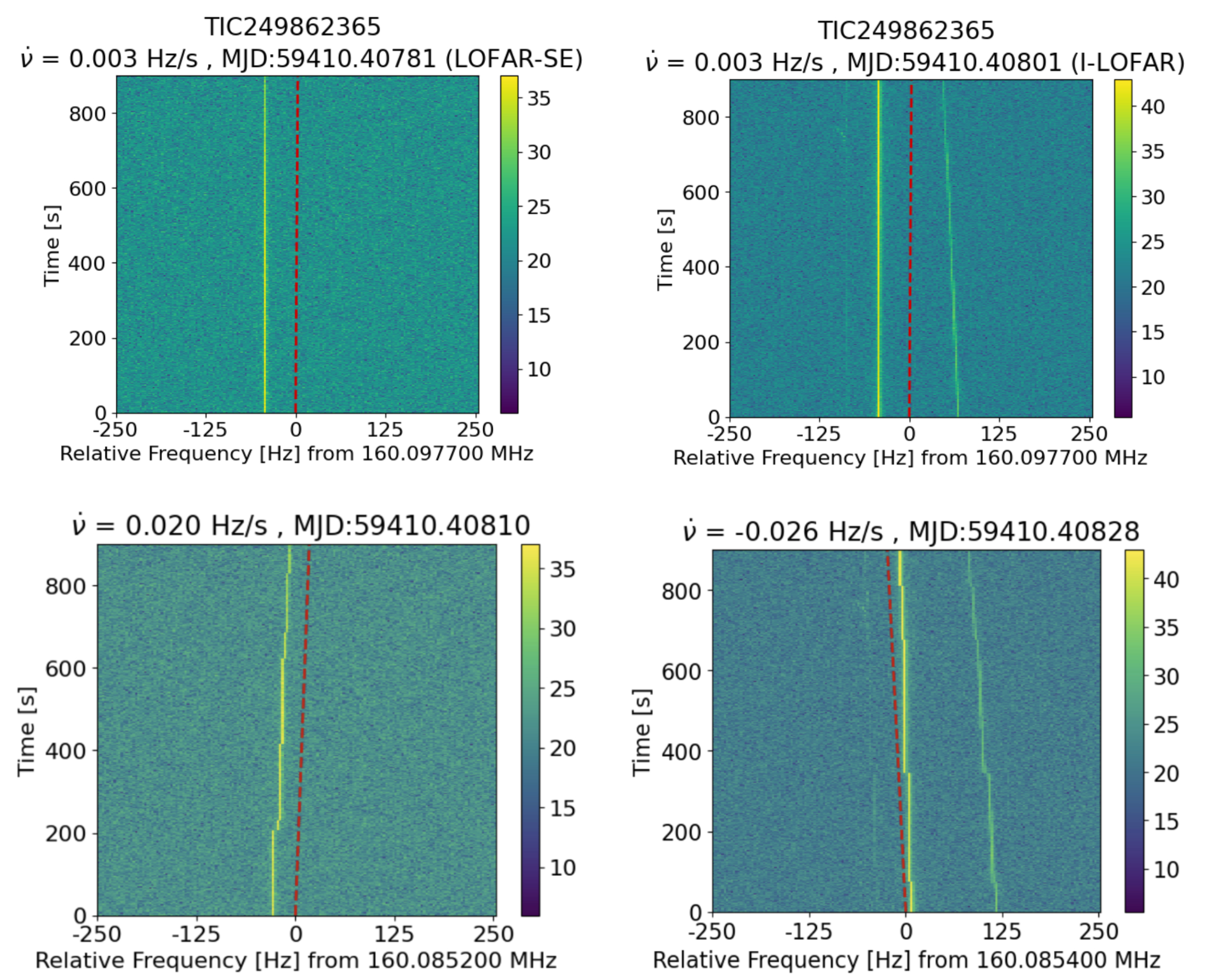}
  \caption{The \textit{top} two plots represent narrowband signals detected at both stations in the topocentric frame of reference. The \textit{bottom} two plots show the same detected narrowband signals detected at both stations but corrected to the barycentric frame of reference. This is illustrated by the newly added drift to signals present in the post correction.}
  \label{fig:barycentric correction}
\end{figure*}

In our analysis, a signal is classified as a mutual extraterrestrial hit only if two conditions are met: \textit{a)} the signals are within a frequency range of $\pm 4 \ \text{Hz}$ of each other in the barycentric reference frame, and \textit{b)} their drift rates are within $\pm 0.2 \ \text{Hz/s}$ of each other after barycentric drift corrections. In Figure \ref{fig:barycentric correction}, an intriguing candidate is depicted. In the topocentric frame, we detected a narrowband signal at 160 MHz that was simultaneously present at both stations. However, when converting to the barycentric reference frame (as illustrated in Figure \ref{fig:barycentric correction}), the signal appears to be seen at different frequencies with opposite signs due to the different line-of-sight velocities towards the target. As a result, this narrowband signal is rejected as a genuine sky-bound signal.

\subsection{Search Results}

\Cref{fig:scatter_DR} depicts the observed drift rates as a function of frequency. As shown in \Cref{fig:histogram_hits_comparison}, a substantial number of hits are observed at 125 MHz and 138 MHz, suggesting the presence of potential aircraft communications within these frequency bands. Notably, at I-LOFAR, a significant number of hits are also detected at 167 MHz, which is suspected to be related to aircraft communication. Additionally, a pronounced spike in RFI at 164 MHz is observed at the Swedish station (LOFAR-SE), likely associated with marine communications due to the proximity of LOFAR-SE to the coast. \\ 

  Implementing barycentric corrections and coincidence rejections, no signals of interest were identified among the observed candidates. Comprehensive information regarding the targets and their corresponding hits for both \textit{TESS} and \textit{Gaia} can be found in the supplementary databases presented in the Tables \ref{tab:TESS_results} and \ref{tab:Gaia_results}.

\begin{deluxetable*}{lCCCCCCC}
  \tablecaption{\textit{TESS} candidates used as bore sight pointings for survey. These targets were obtained from the NASA Exoplanet Archive (NEA) Targets of Interest database. Selection criteria was based on what targets were visible at both stations.   \label{tab:TESS_results}}
  \tablewidth{0pt}
  \tablehead{
  \colhead{TIC ID} & \colhead{MBJD Start} & \colhead{RA} & \colhead{DEC} & \colhead{Distance} & \colhead{LOFAR-SE Hits} &\colhead{I-LOFAR Hits} & \colhead{Mutual Hits}\\   
  \colhead{} & & \colhead{\textit{J2000} Hours} & \colhead{\textit{J2000} Degrees} & \colhead{pc} & & &  \\ \cline{3-5}}
  \startdata
27677846  & 59410.3856 & 4.424676  & 46.365902 & 78.71  \pm 0.330  & 294 & 151 & 0  \\
51024887  & 59410.3189 & 2.810346  & 62.189260 & 41.53  \pm 0.158  & 265 & 181 & 0  \\
81831095  & 59410.3300   & 3.187109  & 61.762385 & 399.40 \pm 5.300  & 277 & 196 & 0  \\
121966220 & 59410.3967 & 5.053188  & 41.785784 & 472.70 \pm 9.135  & 376 & 171 & 0  \\
142090065 & 59402.4266 & 5.271567  & 79.737727 & 182.91 \pm 1.291  & 290 & 282 & 0  \\
191146556 & 59410.4078 & 0.553625  & 46.340305 & 282.83 \pm 3.011  & 378 & 266 & 0  \\
249862365 & 59410.3078 & 2.541534  & 52.704091 & 184.69 \pm 1.290  & 330 & 197 & 0  \\
250724252 & 59410.2967 & 2.233683  & 53.121508 & 574.87 \pm \nodata & 265 & 262 & 0  \\
266500992 & 59410.3634 & 4.078149  & 52.256992 & 165.31 \pm 0.987  & 276 & 164 & 0  \\
288132261 & 59403.5613 & 13.965621 & 79.583322 & 154.94 \pm 0.521  & 328 & 310 & 0  \\ \hline 
  \multicolumn{8}{c}{\textit{First 10 entries of the table, full table available with paper's supplementary material.}} \\ \hline\enddata
  \tablecomments{The above table contains relevant parameters to each observed target and the resultant hits attained from the narrow-band search at respective stations. A * denotes a confirmed exoplanet in the NEA. The distance is inferred from \cite{TIC_Distances} which makes use of GDR2 parallax values. }
\end{deluxetable*}

\begin{deluxetable*}{lCCCCCCCCC}
  \tablecaption{\textit{Gaia} candidates found within $1.295^\circ$ of \textit{TESS} boresight pointing. The \textit{Gaia} target pool makes use of GDR3, filtering was applied to the target pool based on the error of parallax. \textit{Gaia} targets that exhibit a distance error of 20\% or more are cut from the data set as their EIRP sensitivity. \label{tab:Gaia_results}}
  \tablewidth{\textwidth}
  \tabletypesize{\scriptsize}
  \tablehead{
  \colhead{\textit{Gaia} ID} & \colhead{RA} & \colhead{DEC} & \colhead{Distance} & $T_{\text{eff}}$ & \colhead{Pointing Separation} & \colhead{TIC ID} &\colhead{LOFAR-SE Hits} &\colhead{I-LOFAR Hits} & \colhead{Mutual Hits}\\   
  \colhead{} &  \colhead{\textit{J2000} Degrees} & \colhead{\textit{J2000} Degrees} & \colhead{pc} & \colhead{K} & \colhead{deg} & & & &  \\ \cline{2-6}}
  \startdata
540394404387737000 & 0.003867 & 77.7984 & 850.636 & 4954.0 & 1.129 & 407394748 & 257 & 181 & 0 \\ 
540420788369495000 & 0.008171 & 77.9843 & 836.863 & 5387.3 & 1.093 & 407394748 & 257 & 181 & 0 \\ 
540290637978301000 & 0.011819 & 77.5007 & 563.690 & 4686.4 & 1.013 & 407394748 & 257 & 181 & 0 \\ 
540431753423271000 & 0.017921 & 78.2404 & 1293.320 & 5365.0 & 1.295 & 407394748 & 257 & 181 & 0 \\ 
540422957330267000 & 0.018887 & 78.0603 & 722.054 & 5415.4 & 0.958 & 407394748 & 257 & 181 & 0 \\ 
564460068219877000 & 0.018908 & 78.4901 & 1415.490 & 5276.0 & 1.189 & 407394748 & 257 & 181 & 0 \\ 
540394400090438000 & 0.020346 & 77.7933 & 861.226 & 5896.3 & 1.118 & 407394748 & 257 & 181 & 0 \\ 
540387257562774000 & 0.021907 & 77.6559 & 818.213 & 4952.0 & 0.947 & 407394748 & 257 & 181 & 0 \\ 
564451272126871000 & 0.022257 & 78.3080 & 854.196 & 4731.2 & 1.089 & 407394748 & 257 & 181 & 0 \\ 
540291153374376000 & 0.022550 & 77.5313 & 1105.790 & 4892.0 & 0.985 & 407394748 & 257 & 181 & 0 \\ 
\hline   \multicolumn{8}{c}{\textit{First 10 entries of the table, full table available with paper's supplementary material.}} \\ \hline\enddata

 \tablecomments{The above table contains the details of each \textit{Gaia} target found in a specific \textit{TESS} pointing and associated separation from the boar sight of the LOFAR beam.}
\end{deluxetable*}
\section{Discussion}\label{sec:discussion}
\subsection{Survey Sensitivity}
\label{sec:sensitivity}
The required power for a certain extra terrestrial intelligent (ETI) transmitter to be detected depends on its directionality and other signal characteristics. 
We can measure the transmitter power of an ETI beacon in terms of the effective isotropic radiated power (EIRP; \citealt{Enriquez:2017}) as,
\begin{equation}
    {\rm EIRP} = \sigma \times {4 \pi d_\star^2} \;\frac{{\rm SEFD}}{\delta{\nu_t}}\sqrt{\frac{\delta{\nu}}{n_{p}t_{obs}}}  \>\> \mathrm{W}\;
    \label{Eq:EIRP}
\end{equation}
Here, $\sigma$ is the required S/N, $\delta{\nu}$ is the bandwidth of the received signal, 
$\delta{\nu}_{t}$ is the transmitted bandwidth, $t_{obs}$ is the observing integration time, SEFD is the System Equivalent Flux Density, $n_p$ is the number of polarizations, 
and $d_\star$ is the distance between the transmitter and the receiver, i.e., the distance to the star. We considered $\delta{\nu}_{t}$ to be 1\,Hz. For the narrowband signals we consider in our Doppler searches, we assume $\delta{\nu}$ is matched to our spectral resolution and further assume a temporal duty cycle of $100\%$.

The SEFD of the international HBA stations used in this survey is on average $\sim 900$~Jy for most of the band, rising to $\sim 1.2$~kJy at the band edges~\citep{2013arXiv1305.3550V}. For reference, the SEFD of the GBT at the $1.4$~GHz is around 10 Jy. This difference in sensitivity is 
because sky temperature scales as $\nu^{-2.6}$ meaning it can be hundreds of times higher in the LOFAR band. However, the average values are of little use as there is also a large degree of variation in the sky temperature across the sky. For this reason, we perform a separate calculation for every target to determine the relevant luminosity limit in each case.  For example, the sensitivity to each off-boresight target is determined by its proximity to the pointing coordinates. Figure \ref{fig:off_beam_sens} depicts both the impact on sensitivity and the number of background stars within a given LOFAR beam. Figure \ref{fig:inbeam_trgts_SDSS} shows an example pointing from the survey and also well illustrates the vast volume of targets that appear in a LOFAR beam (pink).

\begin{figure*}
  \centering
     \includegraphics[width = \textwidth]{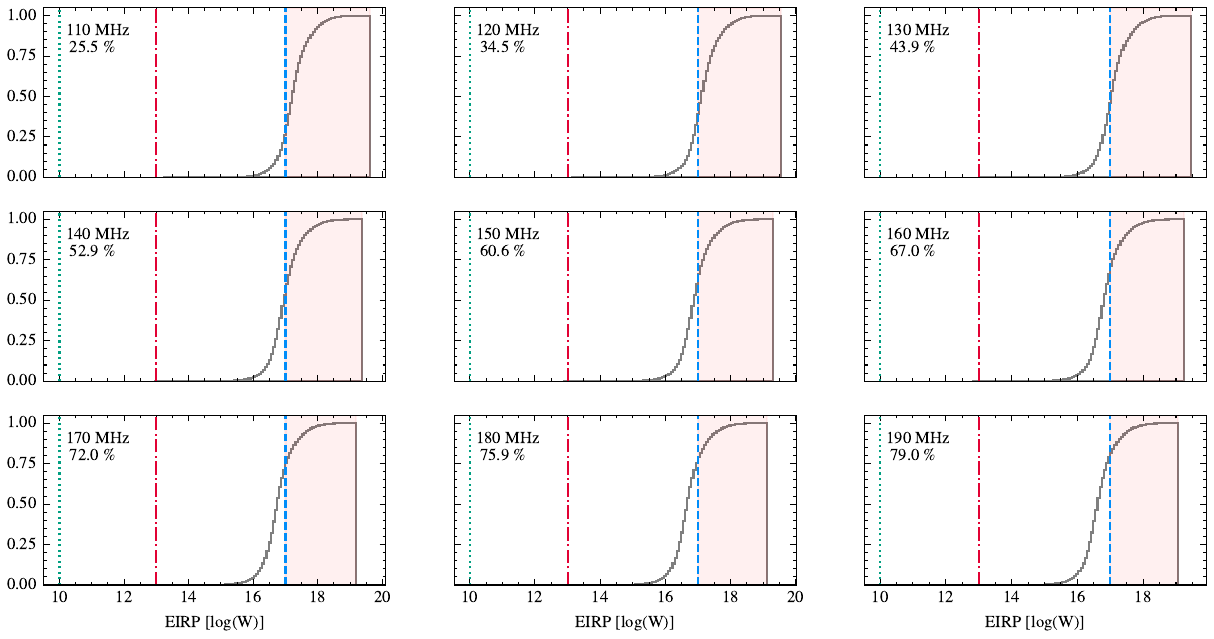}
      \caption{Cumulative histogram of EIRP limits of this survey across the HBA band. Reference luminosities for three civilization levels emitting $10^{17}$, $10^{13}$ or $10^{10}$~W are shown in blue, red and green respectively. The percentage of targets where the station is sensitive to the transmission of $10^{17}$~W is shown, as a function of frequency across the band. At lower frequencies sensitivity to $10^{17}$ emitters drops off as the $T_{sys}$ rises. The $\bar T_{sys}$ varies from 1260 K down to 322 K as frequency is increased across the band. Detailed calculations are presented in Appendix B.}
  \label{fig:Cumulative_EIRP_plots}
\end{figure*}

Figure \ref{fig:Cumulative_EIRP_plots} presents the luminosity limits for the cumulative targets of this survey within the frequency range of $110-190$ MHz. In this figure, the limiting luminosity is compared to notable values of Equivalent Isotropic Radiated Power (EIRP) for various scenarios. These scenarios include a Kardashev I type advanced civilization transmitting at a power level of $10^{17}$ W, an advanced civilization producing planetary radar-level transmissions with a transmitting power of $10^{13}$ W, and a cumulative aircraft radar-type system transmitted across a large solid angle with a power of $10^{10}$ W \citep{Siemion_KEPLER_ApJ}. The Figure demonstrates that due to the varying system temperature ($T_{sys}$) across the frequency band, our observations were sensitive to detecting a range of Kardashev I type targets. Specifically, we were able to detect approximately 25\% of the targets at the lower end of the frequency band, increasing to nearly 80\% of the targets at the higher end of the band.

\subsection{Starlink Interference}
A recent study by \cite{LOFAR-Starlink} has shown that unintended electromagnetic radiation (UEMR) from the Starlink satellite constellation produced broadband interference ranging from 0.1 to 10 Jy and narrowband interference ranging from 10 to 500 Jy. UEMR was detected at frequencies at 110 to 188 MHz, which is the bandwidth of the HBA used in this study. UEMR has many potential consequences for low frequency radio observations. Analysis of \verb|turboSETI| hit detection's within 1 MHz the UEMR narrowband emission frequencies (125, 135, 143, 150, 175 MHz; \citealt{LOFAR-Starlink}) show that 17.8\% of total detected hits are within this region. This region occupies 18.5\% of this study's bandwidth. This result is expected, as UEMR affects a number of different types of radio searches, it does not affect narrowband searches of this nature. This is due to two factors. Firstly satellites in Low Earth Orbit (LEO) to Geostationary orbit (GEO) usually have velocities that are too high for them to be detected with the temporal resolution of our data, and they are at drift rate values outside the parameters of this search ($>$4 Hz/s). Secondly, the satellites are in the near field of each station's beam and due to their relative distance to the observer compared to the bore sight target a satellite less likely to appear in both beams simultaneously. It is concluded that the Starlink constellation does not add to the number of narrowband hits detected in this study.

\subsection{Figure of Merit}
To compare SETI surveys, \cite{Enriquez:2017} introduced a figure of merit known as the Continuous Waveform Transmitter Rate (CWTFM), 

\begin{equation}
    \text{CWTFM} = \zeta_{\text{AO}} \dfrac{\text{EIRP}}{N_{\text{stars}} \nu_{\text{rel}}}
    \label{eq:CWTFM}
\end{equation}

\begin{figure*}
    \centering
    \includegraphics[scale=0.5]{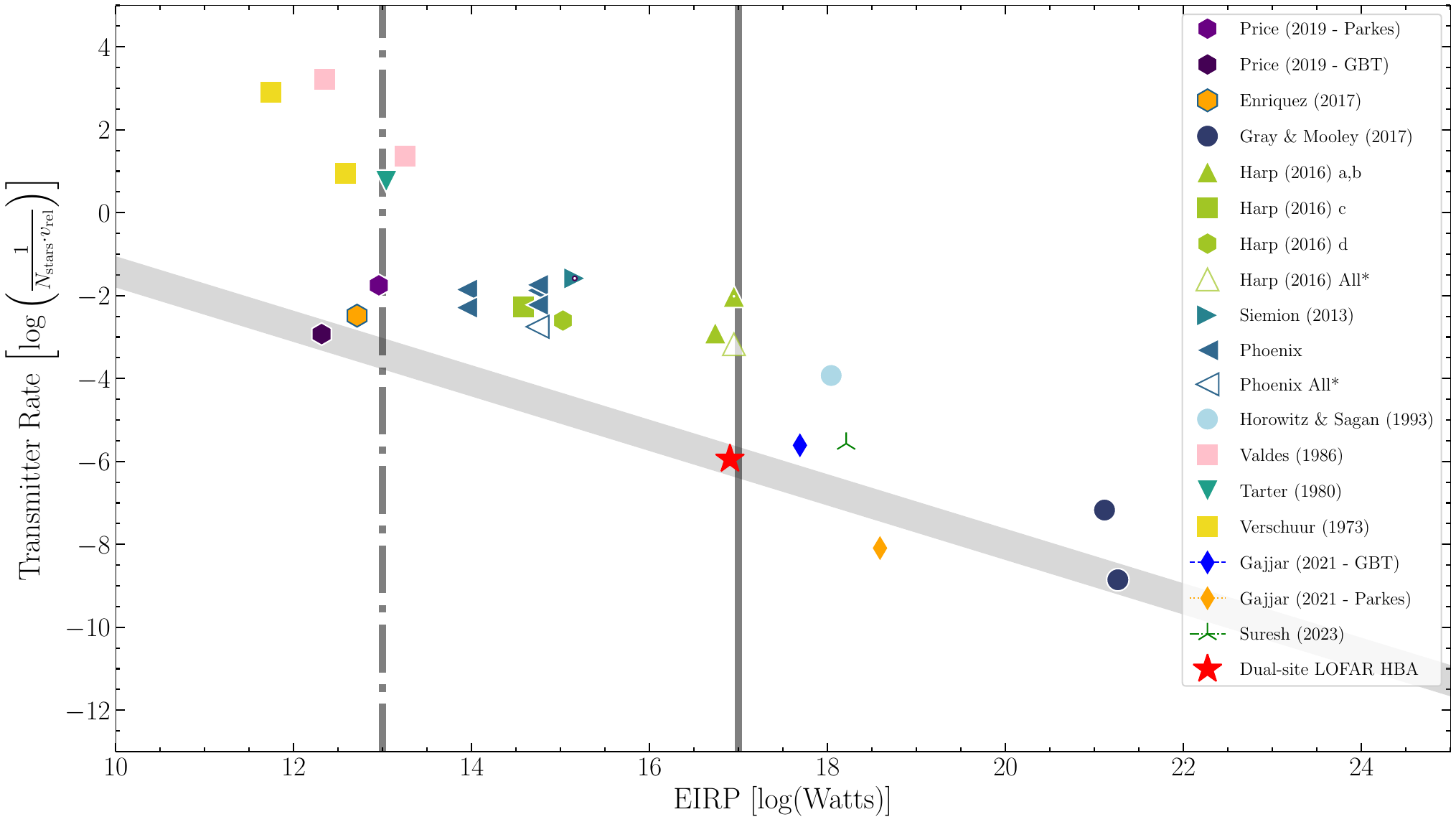}
    \caption{Comaprison of our study (highlighted in red) to prior surveys. The plot presents the transmitter rate versus Effective Isotropic Radiated Power (EIRP), with the grey line indicating the transmitter rate as a function of EIRP. A solid vertical grey line illustrates the energy surplus of a Kardashev Type-I civilization. Additionally, a dotted-dashed grey line depicts the EIRP of the Arecibo planetary radar. The gray thick line shows the slope
    of the transmitter-rate as a function of their EIRP power.  Distance used for EIRP calculation is $\bar d + d_\sigma = 7009 \; \text{ly}$.
    \label{fig:EIRP_compare}}
\end{figure*}

Where $N_{\text{stars}}$ is the number of stars in each pointing for a given survey, $\nu_{\text{rel}}$ is the fractional bandwidth of the survey, $\Delta \nu_{\text{tot}}/\nu_{\text{mid}}$ where $\nu_{\text{mid}}$ is the central frequency of the survey. $\zeta_{\text{AO}}$ is the normalization factor such that $\text{CWTFM} = 1$ and the EIRP is equal to the Arecibo radio telescope's S-band planetary radar, if it were transmitted across a whole hemisphere, $(\sim 10^{13}$~W~\citealt{Siemion_KEPLER_ApJ}). In \Cref{fig:EIRP_compare}, we compare our results to those from some other SETI surveys. As this survey continues, the transmitter rate will decrease with the volume of stars observed $(N_\text{stars})$, as outlined by \Cref{eq:transmitter_rate}.

 \begin{equation}
     \text{Transmitter Rate} = \log \left[ \dfrac{1}{N_{\text{stars}} \cdot \nu_{\text{rel}}} \right]
     \label{eq:transmitter_rate}
 \end{equation}

As dictated by \Cref{Eq:EIRP} to be sensitive to lower powered ETI transmitters, the observation sensitivity needs to be greater. This can be done by employing further LOFAR stations for $n$-site simultaneous observations through coherent summation.
 
\subsection{110 - 190 MHz Technosignature Parameter Space}
 The number of intelligent civilizations is quantified using \Cref{eq:Drake_equation}, coined by \cite{Drake:1961bv}. 
\begin{equation}
    N = R_* f_p N_e f_l f_i f_c L 
    \label{eq:Drake_equation}
\end{equation}
 In this expression $N$ is the number of communicative civilizations in the Milky Way galaxy, $R_{*}$ is the rate of star formation per year in the galaxy, $f_{p}$ is the fraction of stars that have planets around them, $n_{e}$ is the average number of planets that can potentially support life per star that has planets, $f_{l}$ is the fraction of potentially life-supporting planets that actually develop life, $f_{i}$ is the fraction of planets with life that go on to develop intelligent life and $f_{c}$ is the fraction of civilizations that develop a technology that releases detectable signs of their existence into space. \\ 

In \cite{2022VishalPulsed} a modified variation of the Drake equation (Equation. (\ref{eq:modified-drake}); \citealt{Sagan1966}) is used to constrain the fraction of narrowband emitting civilizations $(f_c^n)$.  

\begin{equation}
    N =  R_{\text{IP}}f^n_cL
    \label{eq:modified-drake}
\end{equation}

\begin{figure}
    \centering
    \includegraphics[width = 0.47\textwidth]{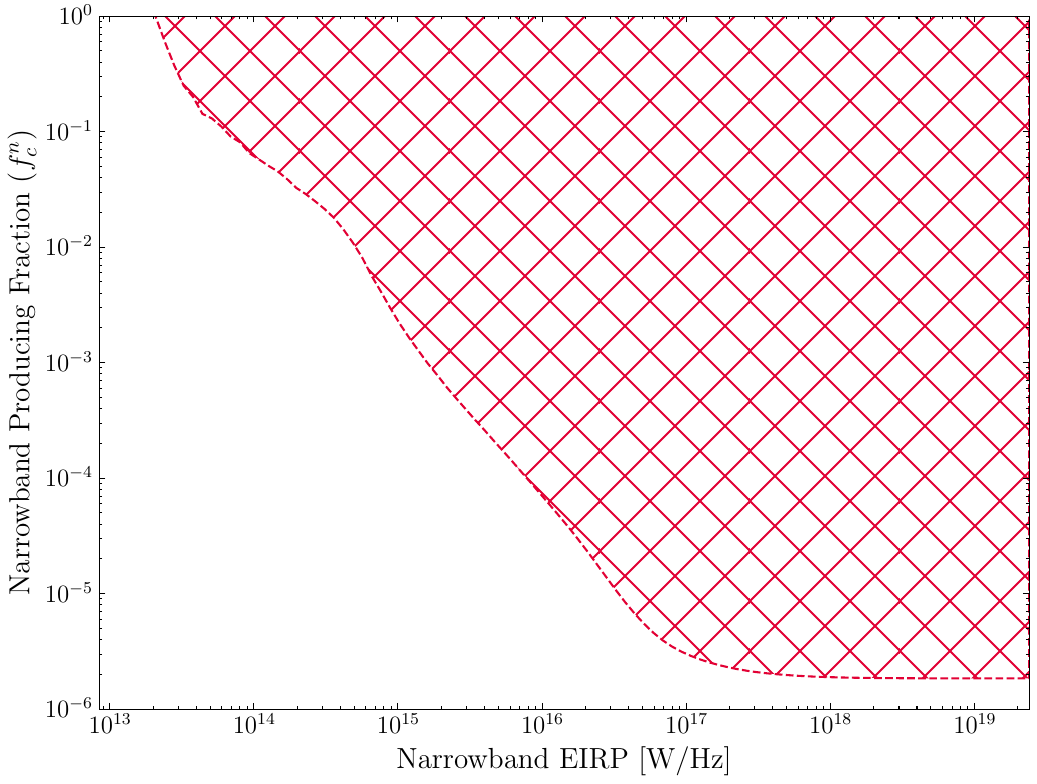}
    \caption{The fraction of stars that produce narrow-band emission ($f^n_c$) against the transmitter power of the total target pool. The hashed region (red) shows the constraints this survey places on a value of $f^n_c$ at 110 - 190 MHz.}
    \label{fig:SETI-constraint}
\end{figure}

Here  $R_{\text{IP}}$ is defined as the emergence rate $(\text{yr}^{-1})$ of intelligent life in the Milky Way.  \Cref{fig:SETI-constraint} shows the constraint that this survey places on $f_c^n$ when using a Poisson sided upper limit at 95\% confidence which in this case is 2.995 as per \cite{1986_Poission_tables}. This provides the most strigent constraint of $f^n_c$ in this frequency range.

\subsection{Future work using LOFAR 2.0}

LOFAR is soon to undergo a staged series of upgrades across all stations in the array. These upgrades at individual stations across Europe will involve the installation of a new Receiver Control Unit (RCU) as described in \cite{LOFAR2}. These RCUs will enable the simultaneous use of both the LBA and HBA in the frequency range of 15 - 240 MHz. This enhancement will allow for a SETI survey across a broader low-frequency band.

Specifically, at 30 MHz, the FWHM will cover an area of 19.39 deg$^2$, decreasing to 1.73 deg$^2$ at 190 MHz. This will enable follow-up LOFAR surveys to encompass a larger volume of stars and a broader frequency domain.
\section{Conclusion}\label{sec:conclusion}

This paper presents a SETI search in a mostly unexplored parameter space as seen with other SETI surveys \citep{Gajjar_2021_BLGC1, Price:2018bv} by simultaneously observing TESS and Gaia targets of interest in the $110-190$~MHz radio window. It also demonstrates dual-site coincidence rejection showing that it provides a new method to discriminate candidate extra-terrestrial signals from terrestrial radio frequency interference. We propose this method as a promising means of follow-up for confirmation of any candidates interest arising in this type of study or others in this frequency range. Each target within our fields was then searched for narrow-band signals at each station using our most up to date search techniques \citep{Enriquez:2017}. The benefit of barycentric correction for eliminating false positives is also demonstrated. Finally the first stringent constraints on the fraction of transmitting civilizations at this frequency range have been shown further contraining the parameter space the Drake equation presents.   

As the LOFAR SETI observation campaign continues and more high resolution frequency data is collected, a machine learning search method comparable to \cite{BLSETI_ML_2023} can be trained and implemented to seek out signals of interest. Multiple LOFAR stations, or indeed sub-arrays of any other wide-footprint radio array allows the option for a coincidence rejection method over the `ON' and `OFF' beam pointings used previously. For future low-frequency SETI surveys, the use of further international stations and a prolonged observation campaign will place even further constraints on an ETI residing in this parameter space. The addition of one or more LOFAR stations would allow for the use of localising a signal of interest in the u-v plane which would be useful for follow up observations. This would be done post-facto through correlation of saved voltages for any candidates of interest.

\section*{ACKNOWLEDGMENTS}
Breakthrough Listen is managed by the Breakthrough Initiatives, sponsored by the
Breakthrough Prize Foundation. D.McK and AB are supported by Government of Ireland Studentships from the Irish Research Council (IRC). I-LOFAR has received funding from Science Foundation Ireland (SFI), the Department of Further and Higher Education, Research, Innovation and Science (DFHERIS) of the Government of Ireland. We acknowledge support from Onsala Space Observatory for the provisioning of its facilities and observational support. 
The Onsala Space Observatory national research infrastructure is funded through Swedish Research Council grant No 2017-00648.
This research has made use of the NASA Exoplanet Archive, which is operated by the California Institute of Technology, under contract with the National Aeronautics and Space Administration under the Exoplanet Exploration Program. This work also presents results from the European Space Agency (ESA) space mission Gaia. Gaia data are being processed by the Gaia Data Processing and Analysis Consortium (DPAC). Funding for the DPAC is provided by national institutions, in particular the institutions participating in the Gaia MultiLateral Agreement (MLA). This work has relied on NASA’s Astrophysics Data System (ADS) Bibliographic Services and the ArXiv. This work has made use of the VizieR catalogue access tool, CDS,
Strasbourg, France.  The original description of the VizieR service was
published in A\&AS 143, 23. We also acknowledge the support and comments from Laura C. Cotter and the Berkeley SETI REU Program, classes of 2021 and 2022. We also thank the anonymous reviewers for their careful reading of our manuscript and their many
insightful comments and suggestions.

\facilities{LOFAR Stations IE613 and SE607}
\software{sigproc \citep{lorimerSIGPROCPulsarSignal2011}, udpPacketManager \citep{David_JOSS}, turboSETI \citep{Enriquez:2017}, blimpy \citep{blimpy}, astropy \citep{astropy}, astroquery \citep{astroquery}, pandas \citep{reback2020pandas}}

\bibliographystyle{aasjournal.bst}
\bibliography{ref} 

\appendix
\counterwithin{figure}{section}
\counterwithin{table}{section}

\section{Post-detection Barycentric Correction} \label{appen:barycenter}
We have developed a novel barycentric correction code specifically designed for high-spectral resolution \texttt{sigproc} filterbank products for technosignature searches \citep{Vishal_Bary}. The code uses the TEMPO routine to calculate relative velocity towards the observing targets at both locations, thus allowing for precise correction of the barycentric drift.

\begin{figure}[h]
    \centering
    \includegraphics[scale=0.45]{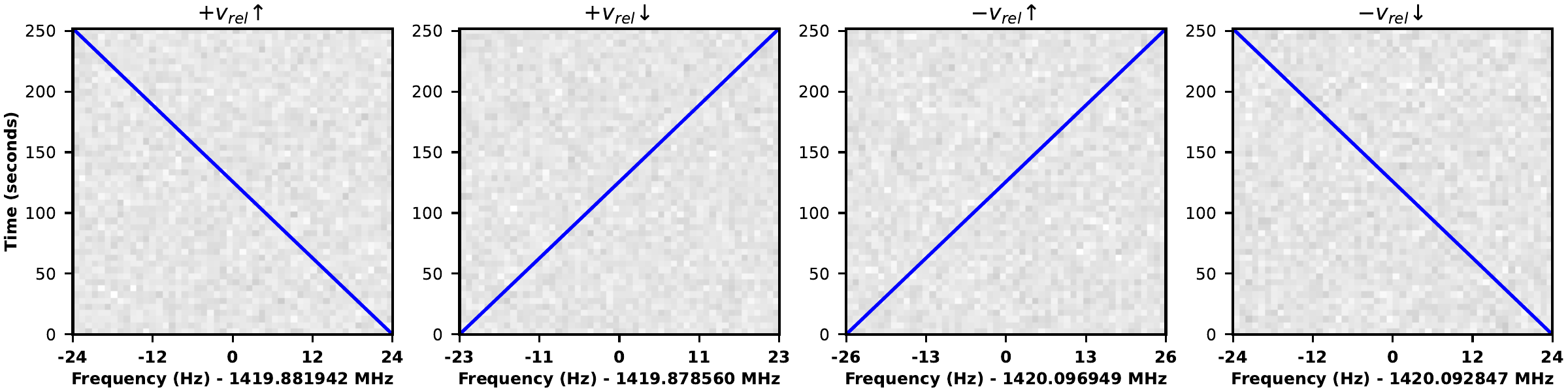}
    \caption{Doppler drift of a narrowband signal in the topocentric observing frame at four different observing epochs. Simulated waterfalls with narrowband signals observed from the Irish LOFAR station towards the direction of TIC\,27677846 are shown for different times of the year in blue. The expected sign and direction of change of the relative velocity are labeled at the top of each plot. It is assumed that a hypothetical narrowband ETI signal is transmitted at a constant frequency of 1420\,MHz (with zero drift rate). As shown, the same signal is observed at different frequencies and drift rates depending on the sign and direction of change of the relative velocity at different epochs of observations. For instance, in the first panel, the relative velocity is positive and increases with observing time. Therefore, the observing frequency has been shifted to a lower frequency (as described by Equation \ref{eq:doppler}), and it continues to shift to even lower frequencies with time.}
    \label{fig:bary_simulated_drift}
\end{figure}
 
The Doppler shift caused by the relative motion between the transmitter and receiver will change over time as it is observed. Over a longer time frame, these shifts will exhibit a sinusoidal curve with a sidereal year period. Over a shorter time frame, the same pattern (superimposed on the yearly pattern) will be visible, but with a sidereal day period. Consequently, the relative velocity will change during the observation period, thereby altering the observed frequency of the received narrowband signal. This leads to a drift in the narrowband signal observed by the observer. Figure \ref{fig:bary_simulated_drift} displays examples of observed drifts at four different epochs for the same narrowband signal source observed from the same location. It is evident that if the relative velocity is positive and increasing with time (leftmost plot in Figure \ref{fig:bary_simulated_drift}), the signal, which is stationary in the barycentric frame, will drift towards lower frequencies as time progresses in the topocentric frame, as per Equation \ref{eq:doppler}.

\subsection{Algorithm outline}
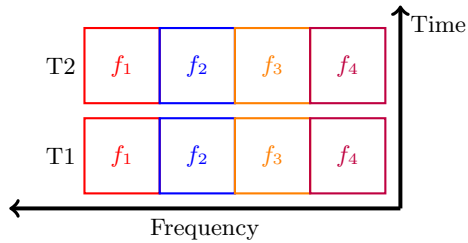
\begin{figure}[h]
\centering
    \begin{tikzpicture}
    \draw[->, line width=1.5pt] (4.2,-0.2) -- (-1,-0.2) node[midway,below] {Frequency};
    \draw[->, line width=1.5pt] (4.2,-0.2) -- (4.2,2.5) node[anchor=north west] {Time};
        \draw[red,thick] (0,1.2) rectangle (1,2.2) node[midway] {$f_{1}$} ;
        \draw[blue,thick] (1,1.2) rectangle (2,2.2) node[midway] {$f_{2}$};
        \draw[orange,thick] (2,1.2) rectangle (3,2.2) node[midway] {$f_{3}$};
        \draw[purple,thick] (3,1.2) rectangle (4,2.2) node[midway] {$f_{4}$};
        \node[anchor=east] at (0,1.7) {T2};
        \draw[red,thick] (0,0) rectangle (1,1) node[midway] {$f_{1}$};
        \draw[blue,thick] (1,0) rectangle (2,1) node[midway] {$f_{2}$};
        \draw[orange,thick] (2,0) rectangle (3,1) node[midway] {$f_{3}$};
        \draw[purple,thick] (3,0) rectangle (4,1) node[midway] {$f_{4}$};
        \node[anchor=east] at (0,0.5) {T1};
    \end{tikzpicture}
    \caption{A typical filterbank file stores data in a time and frequency matrix format. Each row represents a sample, while each column represents a frequency channel for a given sample. Time is increasing from bottom to top while the frequency is increasing from right to left.}
    \label{fig:example_filterbank_file}
\end{figure}

Figure \ref{fig:example_filterbank_file} shows an example of how data is arranged in a filterbank file. For our case, we will assume that in a given filterbank file, frequency channels are in descending order with the first channel ($f_{1}$) having the highest frequency for each time sample. 
The goal of our tool is to shift every frequency channel from the observing frame to the actual emitted frequency frame after correcting for the barycentric relative velocity to remove any additional narrowband signal drift introduce by it. We aim to keep the first channel frequency of all the time samples the same in the barycentric frame also, thus relative shifts between spectras are needed to apply for each time sample corresponding to the inferred relative velocities. Our tool measures relative velocity at each time sample towards a given direction in the sky from a given telescope at the time of observations. Let's assume an observing scenario where $v_{rel}>0$. Following Equation \ref{eq:doppler}, we can state that the emitted frequency (or barycentric frequency) will be higher than the observed frequency. That means that observations stored in the first topocentric frequency channel correspond to higher barycentric frequency. Thus, this spectra needs to be shifted to higher frequency. If the relative velocity increases with time, the consecutive time sample's first channel barycentric frequency will be slightly higher than  the previous sample's first channel emitted frequency. Our tool thus shifts the spectra of each time sample towards higher frequency such that the first channel's emitted frequency matches across all time samples. 
Due to this shifts, we either replace empty channels at the edge of the spectra with zeros in case of squeeze or drop extra channels in case of expansion. 

\begin{figure}[h]
\centering
    \begin{tikzpicture}
    \draw[->, line width=1.5pt] (4.8,-0.2) -- (-1,-0.2) node[midway,below] {Expected Barycentric spectra for $+v_{rel}$};
    \draw[->, line width=1.5pt] (4.8,-0.2) -- (4.8,1.5) node[anchor=north west] {Time};
        \draw[red,thick] (-0.5,0) rectangle (0.5,1) node[midway] {$f_1$};
        \draw[blue,thick] (1.0,0) rectangle (2.0,1) node[midway] {$f_2$};
        \draw[orange,thick] (2.3,0) rectangle (3.3,1) node[midway] {$f_3$};
        \draw[purple,thick] (3.5,0) rectangle (4.5,1) node[midway] {$f_4$};
    \draw[->, line width=1.5pt] (4.8,-3.2) -- (-1,-3.2) node[midway,below] {Corrected Spectra with expansion};
    \draw[->, line width=1.5pt] (4.8,-3.2) -- (4.8,-1.5) node[anchor=north west] {Time};
       \draw[red,dashed] (-0.5,-3) rectangle (0.5,-2) node[midway] {};
        \draw[red,thick] (0.6,-3) rectangle (1.6,-2) node[midway] {$f_1+f_2$};
        \draw[blue,dashed] (1,-3) rectangle (2,-2) node[midway] {};
        \draw[blue,thick] (1.6,-3) rectangle (2.6,-2) node[midway] {$f_2^{'}$};
        \draw[orange,thick] (2.6,-3) rectangle (3.6,-2) node[midway] {$f_3^{'}$};
        \draw[orange,dashed] (2.3,-3) rectangle (3.3,-2) node[midway] {};
        \draw[purple,thick] (3.6,-3) rectangle (4.6,-2) node[midway] {$f_4^{'}$};
        \draw[purple,dashed] (3.5,-3) rectangle (4.5,-2) node[midway] {};
    \end{tikzpicture}
    \begin{tikzpicture}
    \draw[->, line width=1.5pt] (4.8,-0.2) -- (-1,-0.2) node[midway,below] {Expected Barycentric spectra for $-v_{rel}$};
    \draw[->, line width=1.5pt] (4.8,-0.2) -- (4.8,1.5) node[anchor=north west] {Time};
        \draw[red,thick] (1.5,0) rectangle (2.5,1) node[midway] {$f_{1}$};
        \draw[blue,thick] (2,0) rectangle (3,1) node[midway] {$f_{2}$};
        \draw[orange,thick] (2.7,0) rectangle (3.7,1) node[midway] {$f_{3}$};
        \draw[purple,thick] (3.6,0) rectangle (4.6,1) node[midway] {$f_{4}$};
    \draw[->, line width=1.5pt] (4.8,-3.2) -- (-1,-3.2) node[midway,below] {Corrected Frequency channels};
    \draw[->, line width=1.5pt] (4.8,-3.2) -- (4.8,-1.5) node[anchor=north west] {Time};
        \draw[red,thick] (0.6,-3) rectangle (1.6,-2) node[midway] {$f_{1}^{'}$};
        \draw[blue,thick] (1.6,-3) rectangle (2.6,-2) node[midway] {$f_{1}+f_{2}$};
        \draw[orange,thick] (2.6,-3) rectangle (3.6,-2) node[midway] {$f_{3}^{'}$};
        \draw[purple,thick] (3.6,-3) rectangle (4.6,-2) node[midway] {$f_{4}^{'}$};
        \draw[red,dashed] (1.5,-3) rectangle (2.5,-2) node[midway] {};
        \draw[blue,dashed] (2,-3) rectangle (3,-2) node[midway] {};
        \draw[orange,dashed] (2.7,-3) rectangle (3.7,-2) node[midway] {};
        \draw[purple,dashed] (3.6,-3) rectangle (4.6,-2) node[midway] {};
    \end{tikzpicture}
    \caption{These plots depict the expected spectra at their respective barycentric frequencies and the spectra after the correction for barycentric relative velocity. Each plot represents a single spectrum, where the frequency increases from right to left. (a) In the case of $+v_{rel}$, the first channel of the topocentric spectra is shifted to a higher barycentric frequency compared to the last topocentric spectra channel, which causes an expansion of the spectra. If two consecutive channels move farther away from each other (shown as $f_1$ and $f_2$) by more than half the channel width, an additional channel is added in between, which is the summation of these two channels. Extra channels at the edge of the spectra are dropped. (b) In the case of $-v_{rel}$, the first channel of the topocentric frequency is shifted to a lower barycentric frequency relative to the last topocentric spectra channel, which causes a squeeze in the spectra. If two consecutive frequency channels shift closer to each other by more than half a channel width, these channels are added together and an extra channel is added at the end containing zeros. The bottom spectra in each plot illustrate these expansion and squeeze effects after correction are applied from the code. 
    }
    \label{fig:spectra_expands_squeeze}
\end{figure}
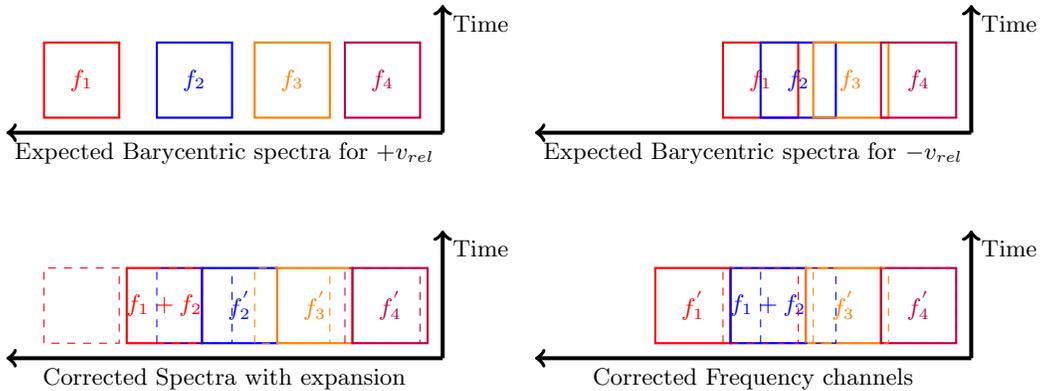

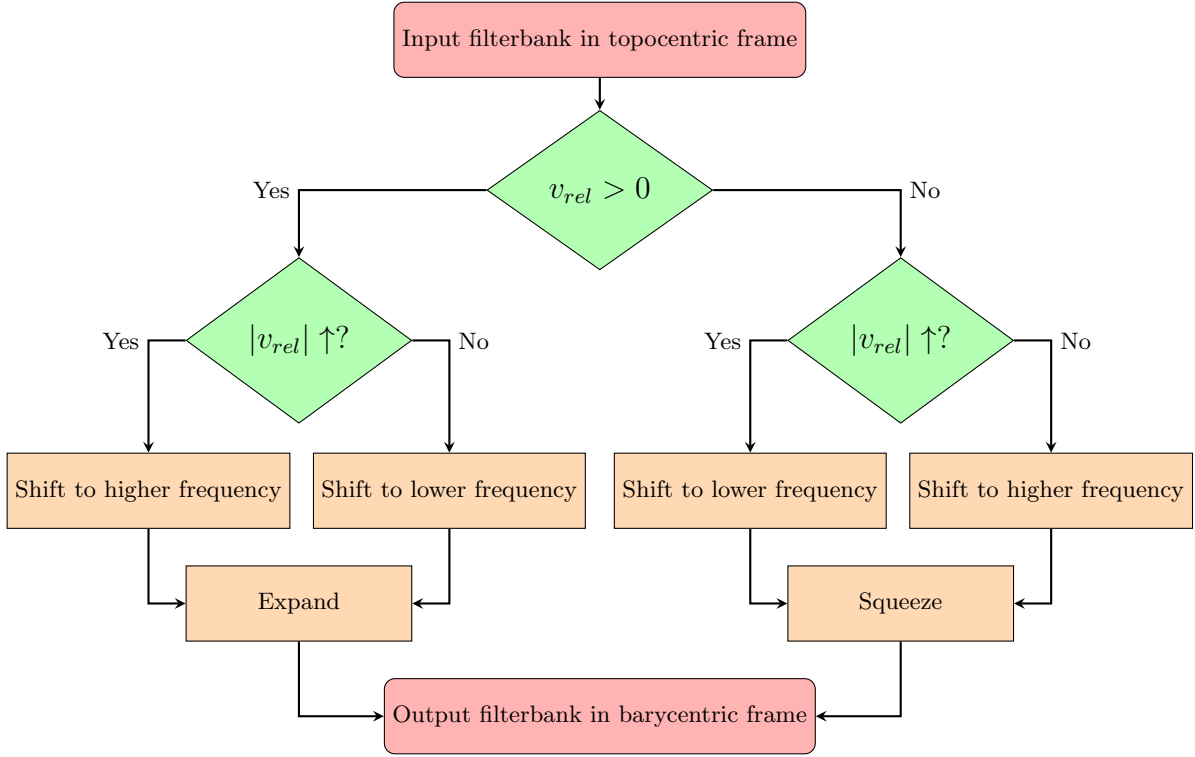
\begin{figure}[h]
    \centering
\begin{tikzpicture}[node distance=2cm]
\tikzstyle{startstop} = [rectangle, rounded corners, minimum width=3cm, minimum height=1cm,text centered, draw=black, fill=red!30]
\tikzstyle{io} = [rectangle, minimum width=3cm, minimum height=1cm, text centered, draw=black, fill=blue!30]
\tikzstyle{process} = [rectangle, minimum width=3cm, minimum height=1cm, text centered, draw=black, fill=orange!30]
\tikzstyle{decision} = [diamond, minimum width=3cm, minimum height=1cm, text centered, draw=black, fill=green!30]
\tikzstyle{arrow} = [thick,->,>=stealth]
    \node (start) [startstop] {Input filterbank in topocentric frame};
    \node (dec1) [decision, below of=start] {\large $v_{rel}>0$};
    \node (dec2) [decision, left of=dec1, xshift=-2cm, yshift=-2cm] {\large $|v_{rel}|\uparrow?$};
    \node (proc1) [process, below of=dec2, xshift=-2cm, yshift=-0.0cm] {Shift to higher frequency};
    \node (proc2) [process, below of=dec2, xshift=2cm, yshift=-0.0cm] {Shift to lower frequency};
    \node (proc3) [process, below of=dec2, yshift=-1.5cm] {Expand};
    \node (dec3) [decision, right of=dec1, xshift=2cm, yshift=-2cm] {\large $|v_{rel}|\uparrow?$};
    \node (proc4) [process, below of=dec3, xshift=-2cm, yshift=-0.0cm] {Shift to lower frequency};
    \node (proc5) [process, below of=dec3, xshift=2cm, yshift=-0.0cm] {Shift to higher frequency};
    \node (proc6) [process, below of=dec3, yshift=-1.5cm] {Squeeze};
    \node (stop) [startstop, below of=dec1, yshift=-5cm] {Output filterbank in barycentric frame};
    \draw [arrow] (start) -- (dec1);
    \draw [arrow] (dec1.west) -| node[anchor=east] {Yes} (dec2.north);
    \draw [arrow] (dec2.west) -| node[anchor=east] {Yes} (proc1.north);
    \draw [arrow] (dec2.east) -| node[anchor=west] {No} (proc2.north);
    \draw [arrow] (proc1.south) |- (proc3.west);
    \draw [arrow] (proc2.south) |- (proc3.east);  
    \draw [arrow] (dec1.east) -| node[anchor=west] {No} (dec3.north);
    \draw [arrow] (dec3.west) -| node[anchor=east] {Yes} (proc4.north);
    \draw [arrow] (dec3.east) -| node[anchor=west] {No} (proc5.north);
    \draw [arrow] (proc4.south) |- (proc6.west);
    \draw [arrow] (proc5.south) |- (proc6.east);
    \draw [arrow] (proc6.south) |- (stop);
    \draw [arrow] (proc3.south) |- (stop);
\end{tikzpicture}
\caption{An outline of the post-detection barycentric correction algorithm for an input filterbank file in \texttt{sigproc}  format. For this case, the input filterbank file has a descending order in frequency, and $v_{rel}$ represents the relative velocity between the transmitter and observer. The algorithm considers two cases depending on whether the relative velocity is positive or negative, which indicates whether the source is moving away from or towards the observer, respectively. Each of these cases is further divided into two where the absolute value of the relative velocity can either increase or decrease, requiring the spectra to be shifted to either the higher or lower frequency end. 
For all cases with $+v_{rel}$, each spectra is expanded, and for $-v_{rel}$, each spectra is squeezed, as shown in Figure \ref{fig:spectra_expands_squeeze}. 
The code then writes each of these spectra into another \texttt{sigproc}  filterbank file, which will have each channel frequency closely corrected to the barycentric frame of reference.}  
\label{fig:code_outline}
\end{figure}

As given in Equation \ref{eq:doppler}, the Doppler shifts are frequency-dependent, impacting higher frequencies more than lower frequencies. In other words, for spectra where frequencies are ordered from higher to lower frequency, the first channel will be shifted more compared to the last frequency channel. To compensate for this, we either expand or squeeze spectra as shown in Figure \ref{fig:spectra_expands_squeeze}. Figure \ref{fig:code_outline} outlines the logical flow of the code for a case of an input filterbank file with a descending order of frequency. By comparing Figure \ref{fig:bary_simulated_drift} and the code outline in Figure \ref{fig:code_outline}, we can consider one of the cases where the relative velocity is negative and increasing in absolute value with time. In this case, we need to shift consecutive spectra to lower and lower frequencies to match up their first frequency channel. Furthermore, for any negative relative velocity (either increasing or decreasing with time), we need to squeeze the individual spectra as shown in Figure \ref{fig:spectra_expands_squeeze}. Similarly, the same can be consider for the case of positive relative velocity. The code for this algorithm is publicly  available here\footnote{\url{https://github.com/gajjarv/BaryCentricCorrection}}.

\section{Sensitivity of the Survey across the HBA}
\label{SEC2:A2}
As previously stated the high band antenna spans from 110 - 190 MHz. The system temperature across the band significantly, which reduces the sensitivity of LOFAR at the lower end of the band-pass (see. \Cref{fig:Cumulative_EIRP_plots}). Calculation of $T_{sys}$ and consquently SEFD follows a modified method as outlined in \S~3.3 of \cite{David-RRAT}. The method differs in the calculation of $T_{sky}$ as this is pointing dependent. This study uses the LWA1 Low Frequency Sky Survey \citep{LWA-survey} for sensitivity analysis. We calculate system equivalent flux density (SEFD; Jy) as follows, 

\begin{equation}
    \text{SEFD} = \frac{2 T_{\text{sys}} k_b}{A_e}
\end{equation}

Where $k_b$ is the Boltzmann constant and $A_e$ is the effective collecting area of a single staion HBA. EIRP is then consequently calculated using \Cref{Eq:EIRP}.

\begin{table}[!ht]
    \centering
    \begin{tabular}{l|ccccccccc}
    \hline
        \textbf{Frequency (MHz)} & 110 & 120 & 130 & 140 & 150 & 160 & 170 & 180 & 190 \\ \hline
        $\bar T_{\text{sys}}$ (K) & 1305.073 &  1051.633 &   862.204 &   717.304 &  604.260 &  514.594 &  442.408 &  383.552 &  335.019 \\
        $\overline{\text{SEFD}}$ (Jy) &2148.125 &  1730.967 &  1419.171 &  1180.668 &  994.601 &  847.011 &  728.195 &  631.320 &  551.435 \\
        $\overline{\text{EIRP}}$ (W) & 17.273 &    17.179 &    17.092 &    17.011 &   16.937 &   16.866 &   16.801 &   16.738 &   16.679 \\ 
        $\text{EIRP}_\text{median}$ (W) & 17.235 &    17.141 &    17.054 &    16.974 &   16.899 &   16.829 &   16.763 &   16.701 &   16.642 \\
         ${\text{EIRP}_\text{max}}$ (W)& 19.628 &    19.537 &    19.454 &    19.377 &    19.304 &   19.237 &   19.173 &   19.112 &   19.055 \\ 
        ${\text{EIRP}_\text{min}}$ (W)&12.627 &    12.627 &    12.627 &    12.627 &   12.627 &   12.627 &   12.627 &   12.627 &   12.627 \\
        K1 detectable (\%) &   25.497 &    34.493 &    43.910 &    52.859 &   60.644 &   66.968 &   71.969 &   75.887 &   79.032 \\
        Earth detectable (\%) & 14.659 &    20.613 &    27.730 &    35.694 &   43.908 &   51.781 &   58.826 &   64.739 &   69.612 \\\hline \hline 
    \end{tabular}
    \caption{Statistics on sensitivity across the HBA band. The K1 and Earth detectable values represent the percentage of the target sample that are detectable.}
\end{table}

\end{document}